\documentclass[twocolumn]{aa} 

\usepackage{natbib}
\usepackage{color}
\usepackage{hyperref}
\usepackage[nolist]{acronym}
\usepackage{verbatim}

\usepackage{graphicx}
\usepackage{txfonts}
\hypersetup{colorlinks=true,allcolors=[rgb]{0,0,0.8}}

\makeatletter
\renewcommand*\aa@pageof{, page \thepage{} of \pageref*{LastPage}}
\makeatother

\begin{document}

   \title{BeyonCE - Light Curve Modelling Beyond Circular Eclipsers}

   \subtitle{I. Shallot Explorer}

   \author{Dirk M. van Dam
          \inst{1}
          \and
          Matthew A. Kenworthy \inst{1}
      }

   \institute{Leiden Observatory, University of Leiden,
   PO Box 9513, 2300 RA Leiden, The Netherlands\\
              \email{dmvandam@strw.leidenuniv.nl}
             }

   \date{Received \today; accepted XXXXX}

 
  \abstract
   {Time-series photometry has given astronomers the tools to study time-dependent astrophysical phenomena, from stellar activity to fast radio bursts and exoplanet transits.
   Transit events in particular are focused primarily on planetary transits, and eclipsing binaries with eclipse geometries that are parameterised with a few variables, while more complex light curves caused by substructure within the transiting object require customized analysis code.
   }
   {We present Beyond Circular Eclipsers (\texttt{BeyonCE}), which reduces the parameter space encompassed by the transit of circum-secondary disc (CSD) systems with azimuthally symmetric non-uniform optical depth profiles.
   By rejecting disc geometries that cannot reproduce the measured gradients within their light curves, we can constrain the size and orientation of discs with complex sub-structure.}
   {We map out all the possible geometries of a disc, calculate the gradients for rings crossing the star, then reject those configurations where the measured gradient of the light curve is greater than the theoretical gradient from the given disc orientation.}
   {We present the fitting code \texttt{BeyonCE} and demonstrate its effectiveness in considerably reducing the parameter space of discs that contain azimuthally symmetric structure by analyzing the light curves seen towards J1407 and PDS 110 which are attributed to CSD transits. }
   {}


   \maketitle
%

\section{Introduction}

Times-series photometry has led to tremendous physical insight into astrophysical processes through the ability to measure the intensity of a star with high cadence, precision and accuracy over long temporal baselines.
Several ground-based surveys i.e., the All Sky Automated Survey for Super-Novae \citep[ASAS-SN,][]{Shappee:etal:2014,Kochanek:etal:2017}, the Asteroid Terrestrial-impact Last Alert System \citep[ATLAS,][]{Tonry:etal:2018,Heinze:etal:2018}, the Super Wide-Angle Search for Planets \citep[SWASP,][]{Pollacco:etal:2006}, and space-based surveys i.e., the \textit{Kepler}\ mission \citep{Kepler}, which was extended to the \textit{K2}\ mission \citep{Howell:etal:2014}, the Transiting Exoplanet Survey Satellite \citep[TESS,][]{TESS:2015}, provide a myriad of data that highlight the vast range of stellar variability exhibited on all time scales.
Intrinsic stellar variability can be caused due to high amplitude optical variability of young stars \citep{Joy:1945}, rotational starspot modulation \citep{Rodono:etal:1986,Olah:etal:1997} or asteroseismology \citep{Handler:2013}. 
Other sources of variability include interactions of the star with other objects or dust orbiting the star.
`Dipper' stars are a class of stars where stellar variability arises due to occultations by dust in the inner boundaries of circumstellar discs which produce transits with depths of up to 50\% \citep{Alencar:etal:2010,Cody:etal:2014,Cody_Hillenbrand:2018}.
\cite{Ansdell:etal:2019a} found that this variability requires the misalignment of the inner protoplanetary disc compared to the circumstellar disc.
Another interesting source of variability is stellar occultations due to the transit of a body.
The transiting bodies could be planets \citep{ProtostarsAndPlanetsBook} characterised by short symmetrical dips with regular periods and fixed transit depths, exo-comets, which are characterised by a saw-tooth eclipse as seen in Figure 2 of \citet{Zieba:etal:2019}, disintegrating planets characterised by regular periods, but varying transit depths due to loss of planetary material \citep{Rappaport:etal:2012,Lieshout:2018,Ridden_Harper:2018}. 
An additional source of variability is the transit of tilted and inclined circum-``planetary'' discs, which due to projection effects, create elliptical occulters that generate asymmetric eclipse profiles in the resultant light curve.
There are two leading theories for gas giant formation.
The first is gravitational instability \citep{Kratter:Lodato:2016}, which is when gravitational instabilities in the protoplanetary disc result in gas and dust collapse under its own gravity and eventually forms a planet with a similar metallicity to the protostellar cloud (the top-down formation).
The second is core-accretion \citep{Helled:etal:2014}, where a core grows to a sufficiently large mass that it starts to pull gas from its surroundings onto itself (bottom-up).
As circumstellar material is transferred into the Hill sphere of the forming protoplanet, the material forms an accretion disc that depletes material on that orbital radius from the star and can produce cavities that are observable in the optical and sub-mm wavelengths \citep{Benisty:etal:2021}, and potentially spiral arms.
In both cases for planet formation a disc forms around the protoplanet, which in turn can potentially support moon-formation \citep{Teachey:Kipping:Schmitt:2018}.
Expectations suggest that these circumplanetary discs are large (order of $\mathrm{AU}$) and should be clearly visible as transits in time-series photometry \citep{Mamajek12}.
Problems arise in the fact that planet-searching algorithms were not designed to identify such transits, as the eclipses would be too long, too deep and too complex to be identified an an exoplanet transit.
The number of parameters to describe the geometry of the transit are already quite large, including the fact that there are many astrophysical effects that need to be taken into consideration.
There are currently several prime candidates for occulting circumplanetary discs: V928 Tau \citep{vanDam:etal:2019}, EPIC 204376071 \citep{Rappaport:etal:2019}, EPIC 220208795 \citep{vanderKamp:2021}.
Some candidates exhibit much more complex sub-structure reminiscent of rings: 1SWASP J140745.93--394542.6J1407 \citep[J1407,][]{Kenworthy_2015} and PDS 110 \citep{Osborn2017,Osborn:etal:2019}.

The Shallot Explorer is a module of the \texttt{BeyonCE} (Beyond Common Eclipsers) package that can limit the extended parameter space of circumplanetary ring systems (Section~\ref{sec:parameters}), based on measurements taken from the light curve (Section~\ref{sec:lightcurvedata}). 
The Shallot Explorer is described in detail in Section~\ref{sec:shallot}, with the results of simulations described in Section~\ref{sec:simulations} and validations of the parameter space on real data will be discussed in Section~\ref{sec:validations}. 
Finally Section~\ref{sec:discussionandconclusions} presents the discussion and conclusions.

\section{Circum-secondary Disc Parameters}
\label{sec:parameters}

\begin{figure}
   \centering
   \includegraphics[width=\hsize]{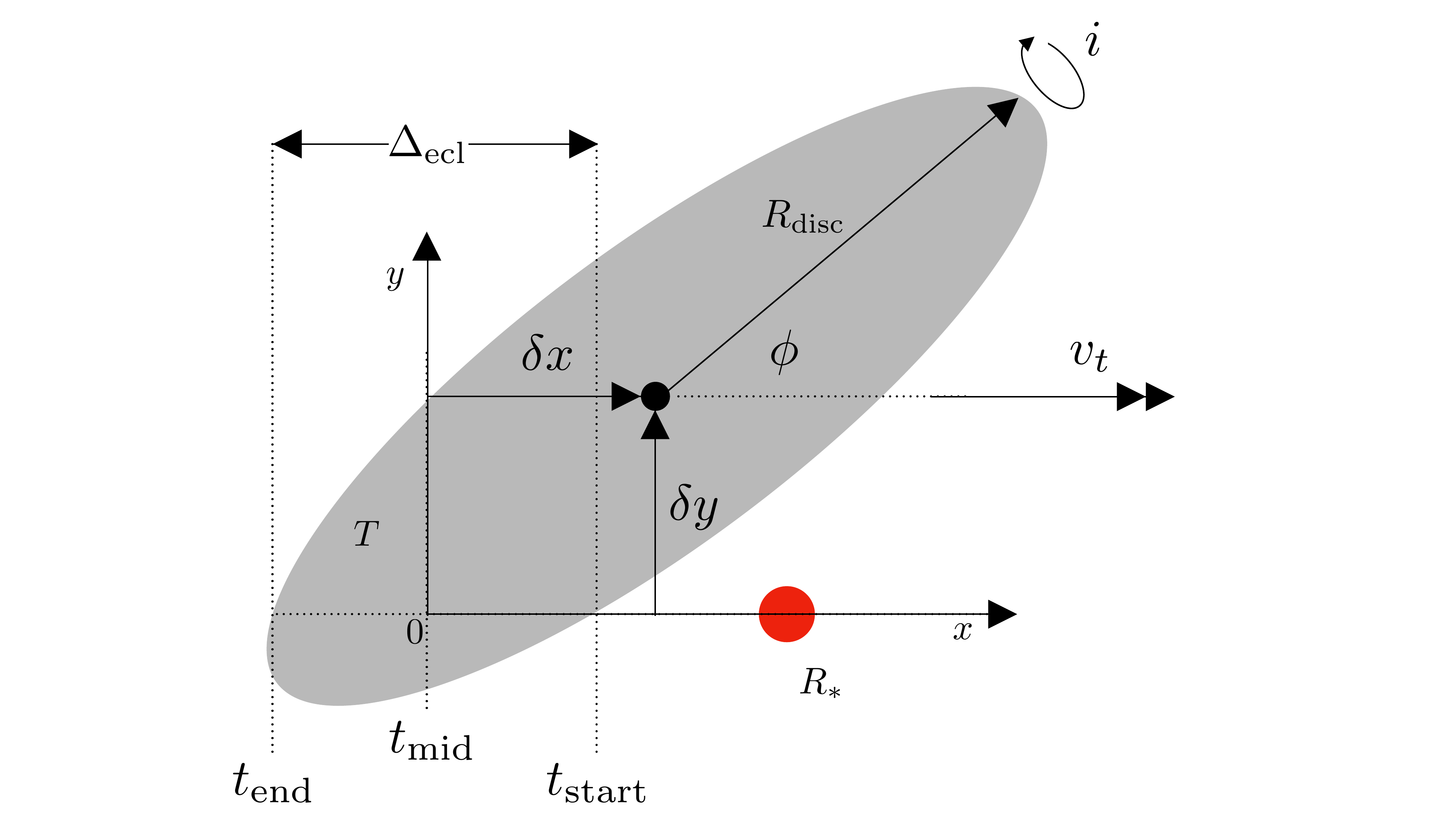}
      \caption{Disc Parameters. See Table~\ref{tab:parameters} to relate the symbols to the parameters. Scale factors in the table are not included here as they are contained by $R_\mathrm{disc}$ and $i$.}
         \label{fig:parameters}
\end{figure}

The parameters that define the geometry and orientation of the disc with respect to our line of sight are shown in Figure~\ref{fig:parameters} and listed in Table~\ref{tab:parameters}.

\begin{table}
    \caption{Disc Parameters.}
    \label{tab:parameters}
    \centering
    \begin{tabular}{l c c c}
        \hline\hline\\
         Parameter & Symbol & Unit & Range \\
        \hline\\
         Disc Radius                    & $R_\mathrm{disc}$    & day                   & $ > 0$       \\
         Transmission                   & $T$           & \textendash                  & 0 \textendash 1\\
         Impact Parameter               & $\delta y$    & day                          & $ > 0$\\
         Centroid Shift                 & $\delta x$    & day                          & \textendash \\
         Tilt                           & $\phi$        & deg                          & 0 \textendash  180 \\
         Inclination                    & $i$           & deg                          & 0 \textendash  90 \\
         Transverse Velocity            & $v_t$         & $R_* \, \mathrm{day}^{-1}$   & $ > 0$\\\\
        \hline\\
         Horizontal Scale Factor        & $f_x$         & \textendash                  & $ > 0$ \\
         Vertical Scale Factor          & $f_y$         & \textendash                  & $ > 0$ \\
        \hline
    \end{tabular}
    \tablefoot{See Figure~\ref{fig:parameters} for a visual representation of these parameters. The scale factors are not included in the figure as they are contained by $R_\mathrm{disc}$ and $i$.}
\end{table}

The disc is assumed to be circular, with a scale height ($h_\mathrm{disc}$) considerably smaller than the radius of the star ($R_*$) and the radius of the disc ($R_\mathrm{disc}$).

$$h_\mathrm{disc} << R_* << R_\mathrm{disc}$$

The disc is inclined to our line of sight by angle $i$ where $i = 0^\circ$ is a face-on disc and $i = 90^\circ$ is an edge-on disc.
This presents an elliptical cross section with semi-major axis of $R_\mathrm{disc}$ and semi-minor axis of $R_\mathrm{disc}\cos i$.
The path of the disc in front of the star with radius of $R_*$ is defined to be in a straight line moving with a constant projected transverse velocity of $v_t$.
The angle between the path of the disc and the on-sky projected semi-major axis of the disc is defined as the tilt $\phi$.
The perpendicular distance (w.r.t. the path of the disc) from the centre of the projected disc to the centre of the star is the impact parameter $\delta y$.
Since the diameter of the star typically contains the largest relative uncertainty of all the physical components in the system, we choose the units of the size of the stellar disc in units of time, usually days, where the physical size of the disc can be converted from time to distance by multiplying by $v_t$.

The inclination, tilt, and radius of the disc define the geometry of the disc projected onto the plane of the sky.
The light curve is characterised by the first ingress of the star at $t_\mathrm{start}$ and egress at $t_\mathrm{end}$.
Exactly between these two values lies $t_\mathrm{mid}$, such that $t_\mathrm{mid} - t_\mathrm{start} = t_\mathrm{end} - t_\mathrm{mid}$.
The disc geometry combined with the impact parameter $\delta y$ therefore defines the chord that the disc makes as it moves in front of the star.
We limit the range of $\delta y$ to be a positive value and the $0 \le \phi \le 180$ because there is a geometric degeneracy in the transit geometry.
The transit geometry for an occulter with $\delta y = y_0$ and $\phi = \phi_0$ is the same as an occulter with $\delta_y = - y_0$ and $\phi = 180 - \phi_0$.

We define a Cartesian coordinate system in the plane of the sky, with the origin centred at the midpoint of the eclipse, $t_\mathrm{mid} = 0$. 
The centre of the projected disc is at $(\delta x,\, \delta y)$ as shown in Figure~\ref{fig:parameters}. 

Though there are many effects that alter the evolution of a light curve during an eclipse such as intrinsic stellar variability, forward scattering of light from the star on the circum-``planetary'' disc \citep[see][]{vanKooten:Kenworthy:Doelman:2020,Budaj:2013}, magnetic interactions between the circumstellar material and the host star \citep{Kennedy:etal:2017}; we consider the limiting case where a star can be modelled as a sphere, with radius, $R_*$.
We allow the model of the star to include limb-darkening as described by the linear limb-darkening law, which is parameterised by the limb-darkening parameter, $u$, as described in Equation~\ref{eq:linearlimbdarkening}. 
\begin{equation}
    \label{eq:linearlimbdarkening}
    \frac{I(\mu)}{I(1)} = 1 - u\,(1 - \mu)
\end{equation}
Note that $\mu = \cos{\gamma}$, where $\gamma$ is the angle between the line of sight and the emergent intensity and $I(1)$ is the specific intensity at the centre of the stellar disc.

The occulter itself can be separated into various parameters.
Starting with the physical objects, you have the anchor body itself, which we assume to be a fully opaque ($T = 0$) spherical body, with radius, $R_p$.
This is supplemented by a larger disc centered on the body, which has a radius $R_\mathrm{disc}$.
This disc can have an intricate radial evolution of optical depth (this includes gaps, i.e. rings). 
This is modelled by separating the disc into any number of rings with an inner radius, $R_{x,\mathrm{in}}$, an outer radius, $R_{x,\mathrm{out}}$, and a ring transmission $T_x$ that varies as a function of radial distance from the central body.
The final parameter for the system is the centroid shift, $\delta x$, which is necessary to define the centre of the occulter w.r.t. the centre of the star.

The model simplifications are reiterated here. 
First, the stellar limb is assumed to be spherical, the star quiescent with no significant star spots on its disc, and with a limb-darkening profile that can be described by a linear limb-darkening law (Equation~\ref{eq:linearlimbdarkening}).
The disc is assumed to be azimuthally symmetric about the anchor body both in geometry and transmission. 
This means that we assume that the transmission profile is one dimensional, $T = T(R)$, and that there are no disc misalignments ($i \neq i(R)$ and $\phi \neq \phi(R)$).
Finally, we assume a constant velocity and impact parameter for the duration of the eclipse (i.e., the path of the occulter as it transits the star is straight).
This assumption is a consequence of the $R_\mathrm{disc} >> R_*$ and is justified in Appendix~\ref{sec:derivation}.

\section{Light Curve Measurements}
\label{sec:lightcurvedata}

A light curve provides information about the star and the transiting object.
We can determine the eclipse duration, which relates to the orbital period and transverse velocity of the object.
We can measure the eclipse depth, which gives an indication of the size (convolved with the transmission) of the occulting object.
If the eclipse exhibits a flat bottom, then the amount of light the occulter is blocking does not change with time, which can mean that either the object is fully contained by the star (as with a planet transit), or the object is completely covering the disc of the star.
The light curve gradients ($dI/dt$) in the light curve provide information about the speed of the occulting object and, if the object is large enough, can also provide information about the geometric tangent of the occulter w.r.t. the host star.
\citet{vanWerkhoven:etal:2014} determined that a lower bound on the velocity of the transiting object could be determined by measuring the steepest light curve gradient in the light curve - the slowest moving occulter that can produce the given eclipse would be a completely opaque boundary large enough to show no curvature travelling at some velocity.
This is because a curved boundary would occult a smaller area of the stellar limb than a straight boundary at each position, requiring a higher velocity of the occulter.
Likewise a lower opacity of the body would require the object to occult a larger area of the star more quickly than a fully opaque object.
The lower bound on the transverse velocity thus depends on the limb-darkening of the star and the steepest light curve gradient of the system, which can be written as Equation~\ref{eq:vanwerkhoven}, where $\dot{L}$ is the steepest light curve gradient in normalised luminosity (using $\dot{L} \left[\mathrm{day}^{-1}\right]$ gives $v_t$ in $R_* \, \mathrm{day}^{-1}$).

\begin{equation}
    \label{eq:vanwerkhoven}
    v_t = \pi \dot{L} R_* \left(\frac{2 u - 6}{12 - 12 u + 3 \pi u}\right)
\end{equation}

The caveat w.r.t. curved boundaries always eclipsing a smaller area of the star is good enough to first order as the Shallot Explorer is not a disc fitter.
Instead it is a useful tool to limit the parameter space that needs to be explored in more detail, where the curved ring boundaries can subsequently be analyzed completely.

\section{Shallot Explorer}
\label{sec:shallot}

The Shallot Explorer is used to describe the parameter space spanned by eclipsing inclined and tilted discs and narrow down the possible models of the disc for further analysis.
Initially it is important that the parameter space described in Section~\ref{sec:parameters} can be described in a grid that can be explored with ease .
For that purpose, the argument is made that when analysing light curves and modelling occulters, the starting point is to determine what the smallest possible disc could be that could cause the eclipses observed.
We are aware that the prior distribution may be flat towards the Hill sphere, but we choose to start with smaller discs due to stability considerations.
The arguments here are made because smaller discs are more likely to be stable as the stabilising influence of the anchor body is stronger for smaller discs and thus dominates over the disturbing influence of the host star and other gravitational interactions \citep{Zanazzi:Lai:2017}.
A light curve produces one very hard limit and that is the duration of the eclipse, which is obtained by converting the combination of the width of the occulting object in the transit chord and the diameter of the star to a time using the transverse velocity.
For the limiting case of $R_\mathrm{disc} >> R_*$ the diameter of the star can be ignored, which allows disc size modelling to be done analytically.

The maths, that will be described in the following sections, show that the parameter grid can linearly scale with the duration of the eclipse.
Thus, the grid is to be defined in terms of eclipse duration, as it can subsequently be transformed to physical scales with the transverse velocity.
This linear scaling encourages the preparation of a high-resolution grid that can be either refined at the appropriate location or linearly interpolated for a more precise investigation.

\subsection{Simple Sheared Disc Model} 
\label{subsec:smalldisc}

To explore this grid in an intuitive manner, we define a Cartesian coordinate system, in eclipse duration space, centred on the midpoint of the eclipse ($t_\mathrm{mid} = 0$), with the $x$-axis aligned with the transit path of the star, and the $y$-axis with the impact parameter.
Note that since we are proposing that the transverse velocity is positive and move in the $x$ direction, larger $x$ point to earlier times in the light curve.
We thus want to find the simplest sheared disc that is bound by the transit duration, $\Delta_\mathrm{ecl}$ and is centred at $\left(\delta x, \, \delta y \right)$.

\begin{figure}
   \centering
   \includegraphics[width=\hsize]{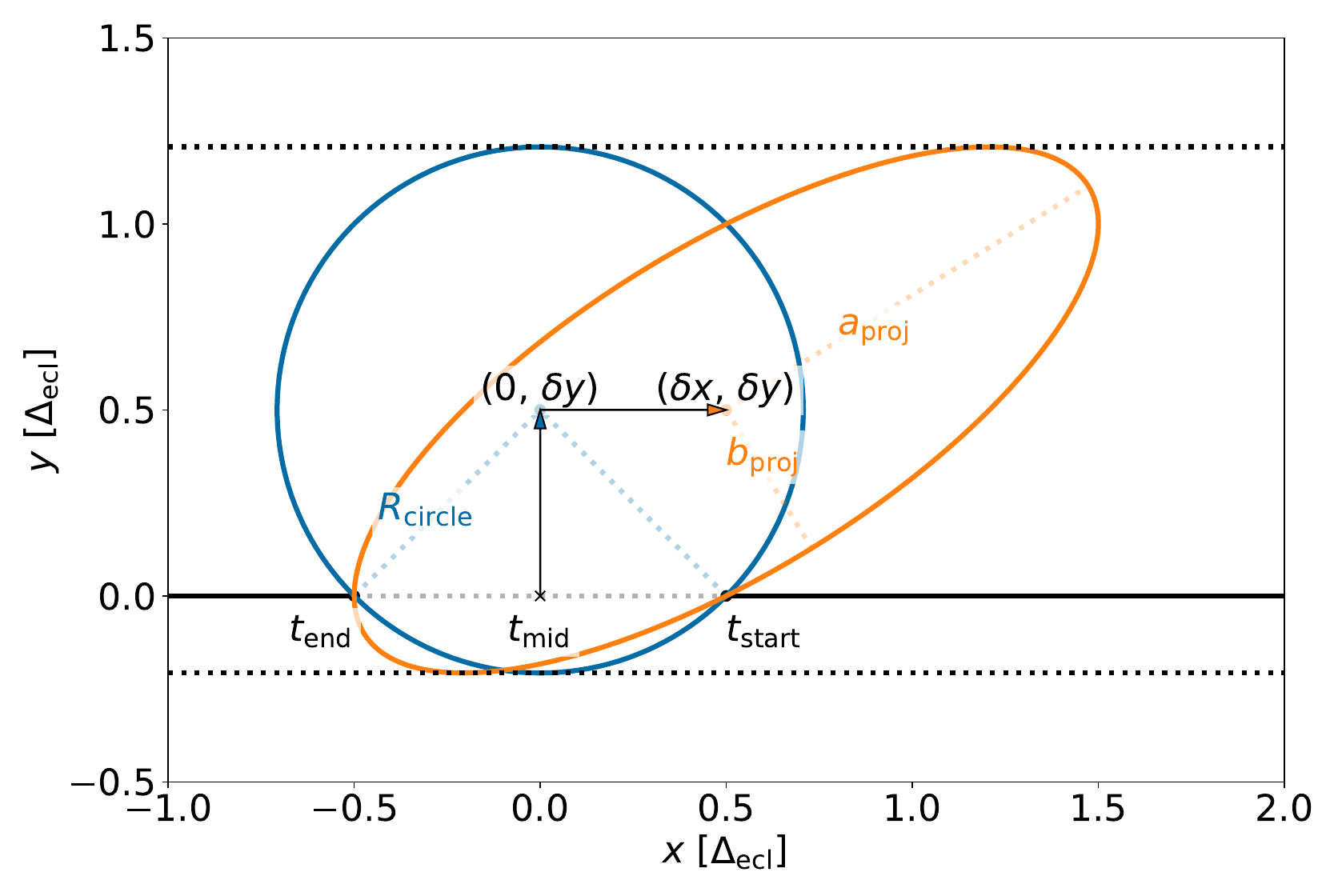}
      \caption{Determining the simplest sheared disc model.
      The coordinate system is centred at $(0, 0)$.
      The midpoint of the disc is shifted up to $(0, \delta y)$.
      A circle (blue) is drawn with radius $R_\mathrm{circle}$ (Equation~\ref{eq:minradiusdisc}).
      The midpoint of the disc is shifted to $(\delta x, \delta y)$, shearing the circle into an ellipse (orange, Equation~\ref{eq:xshear},~\ref{eq:yshear} and~\ref{eq:shear}).
      From this ellipse the disc parameters can be determined.}
         \label{fig:shallotshearing}
\end{figure}

The most obvious solution is simply a circle with $R_\mathrm{disc} = 0.5\, \Delta_\mathrm{ecl}$ centred at $\left(\delta x, \, \delta y \right) = \left(0, \, 0\right)$.
Changing the impact parameter impacts the $y$-coordinate of the midpoint of the disc, which changes the size of the circular disc to a radius defined by Equation~\ref{eq:minradiusdisc} and as seen in Figure~\ref{fig:shallotshearing}.
\begin{equation}
    \label{eq:minradiusdisc}
    R_\mathrm{circle} = R_\mathrm{disc, \, \delta x = 0} = \sqrt{\left(\delta y\right)^2 + \left(\frac{1}{2}\right)^2}
\end{equation}
If we also move the centre of the disc in the $x$-direction, we need to ensure that the resulting ellipse is still bound by $t_\mathrm{start}$ and $t_\mathrm{end}$ at $y = 0$.
To do this we shear the circle into an ellipse, with the transformation defined in Equation~\ref{eq:xshear} and~\ref{eq:yshear}, where $s$ is the shear parameter defined in Equation~\ref{eq:shear}.
\begin{equation}
    \label{eq:xshear}
    x' = x - s\,y + \delta x
\end{equation}
\begin{equation}
    \label{eq:yshear}
    y' = y + \delta y
\end{equation}
\begin{equation}
    \label{eq:shear}
    s = - \frac{\delta x}{\delta y}
\end{equation}
This shear transformation allows the determination of the simplest possible ellipse produced from the sheared circle centred at ($\delta x$, \, $\delta y$), that has a width of $\Delta_\mathrm{ecl}$ and passes through $t_\mathrm{start}$ and $t_\mathrm{end}$ at a height of $y = 0$.
For this grid point the ellipse parameters must be determined, which can in turn be mapped to the disc parameters.
To determine the semi-major, $a_\mathrm{proj}$, and semi-minor, $b_\mathrm{proj}$, axes of the projected disc one must determine the location of a vertex and co-vertex of the ellipse described.
This can be done by noting that the vertex of an ellipse must fulfill the condition $dR/d\theta = 0$, where $\theta$ is the parametric angle of the ellipse.
To perform this operation, we initially define the simple radius circle for the ellipse (see blue circle in Figure~\ref{fig:shallotshearing}) in parametric form in Equations~\ref{eq:circlexparametric} and~\ref{eq:circleyparametric}.
\begin{equation}
    \label{eq:circlexparametric}
    x = R_\mathrm{circle} \cos(\theta)
\end{equation}
\begin{equation}
    \label{eq:circleyparametric}
     y = R_\mathrm{circle} \sin(\theta)
\end{equation}
We then apply the shearing as defined by Equations~\ref{eq:xshear} and~\ref{eq:yshear}.
\begin{equation}
    \label{eq:ellipsexparameteric}
    x' = R_\mathrm{circle} \cos(\theta) - s R_\mathrm{circle} \sin(\theta) + \delta x \\
\end{equation}
\begin{equation}
    \label{eq:ellipseyparametric}
    y' = R_\mathrm{circle} \sin(\theta) + \delta y
\end{equation}
The equation of the ellipse can be written as:
\begin{equation}
    \label{eq:rellipse}
    R'^2 = x'^2 + y'^2
\end{equation}
Substituting Equations~\ref{eq:ellipsexparameteric} and~\ref{eq:ellipseyparametric} into Equation~\ref{eq:rellipse}, taking the derivative and setting it to 0 ($dR/d\theta = 0$) provides the analytic definition (Equation~\ref{eq:thetamaxmin}) of a vertex of this ellipse.
\begin{equation}
    \label{eq:thetamaxmin}
    \theta = \frac{1}{2} \arctan\left(\frac{2}{s}\right)
\end{equation}
This can, in turn be used with Equations~\ref{eq:circlexparametric} and~\ref{eq:circleyparametric} to determine the $(x, \, y)$ coordinate of either $a_\mathrm{proj}$ or $b_\mathrm{proj}$.
The next vertex (either $a_\mathrm{proj}$ if the angle defined by Equation~\ref{eq:thetamaxmin} defines $b_\mathrm{proj}$ or vice versa) is at $\theta + \pi/2$.
With the $(x,y)$ coordinates of the vertices of the ellipse, one can determine $a_\mathrm{proj}$ and $b_\mathrm{proj}$ simply, using Equation~\ref{eq:xytoab}.
\begin{equation}
    \label{eq:xytoab}
    (a,\,b)_\mathrm{disc} = \sqrt{(x_{(a,\,b)} - \delta x)^2 + (y_{(a,\,b)} - \delta y)^2}
\end{equation}
The tilt is obtained by choosing the vertex related to the $a_\mathrm{proj}$ and using Equation~\ref{eq:tilt}.
\begin{equation}
    \label{eq:tilt}
    \phi = \arctan \left( \frac{y_a - \delta y}{x_a - \delta x} \right)
\end{equation}
Finally the inclination is determined by Equation~\ref{eq:inclination}.
\begin{equation}
    \label{eq:inclination}
    i = \arccos\left(\frac{b_\mathrm{proj}}{a_\mathrm{proj}}\right)
\end{equation}

With this process it is possible to determine the simplest sheared disc that has its midpoint at $(\delta x,\delta y)$ and passes through the points $(\pm 0.5, 0)$.
We are able to determine the size and orientation of the simplest disc that could cause an eclipse lasting $\Delta_\mathrm{ecl}$ given the centre of the disc.
Note that in units of time, we need not concern ourselves with $v_t$ as this is simply a proportionality factor to convert from sizes in time to sizes in distance.

\subsection{Extension of Discs}
\label{subsec:largedisc}

\begin{figure}
   \centering
   \includegraphics[width=\hsize]{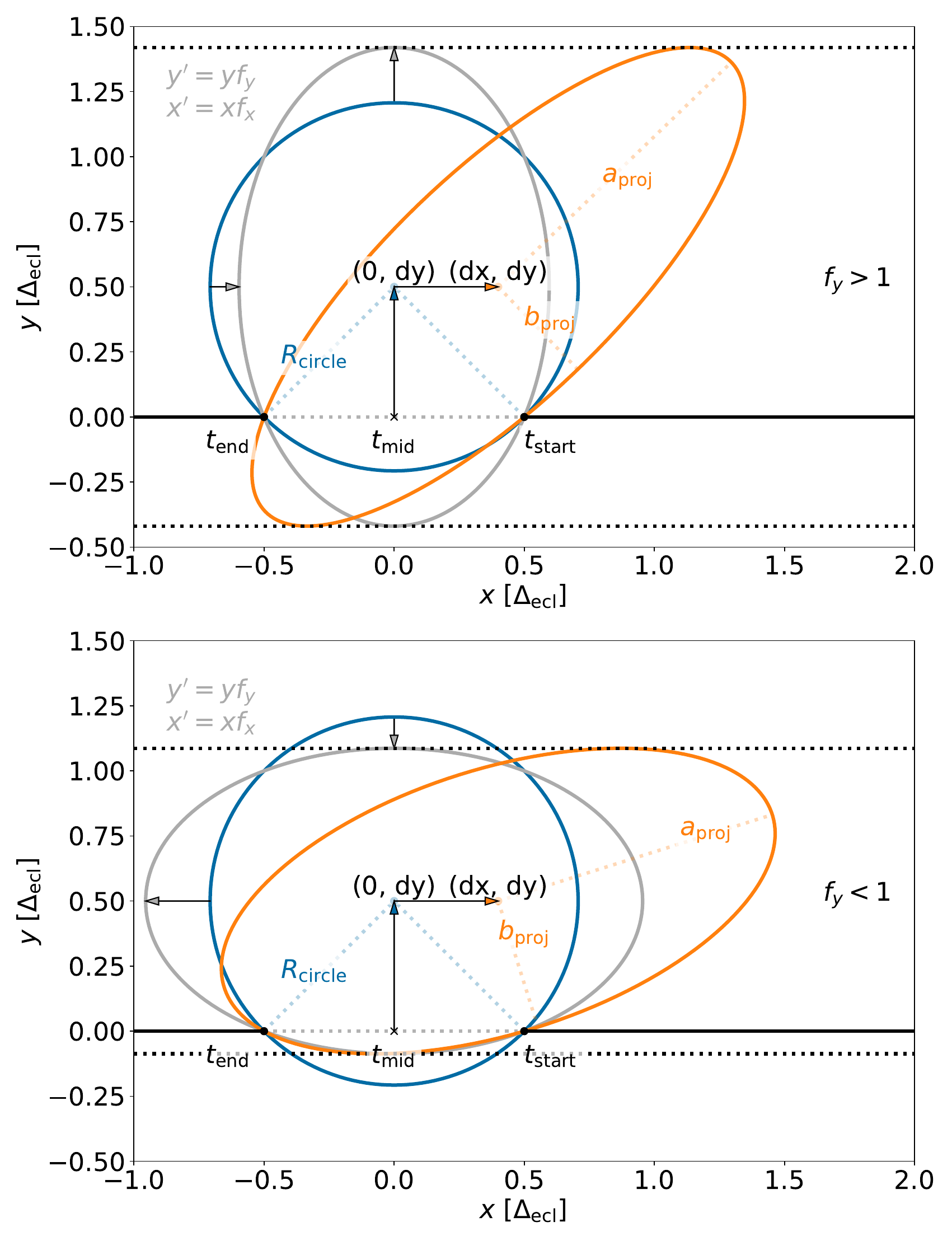}
      \caption{Determining the Scaled Disc Model.
      \textit{Top:} $f_y > 1$. \textit{Bottom:} $f_y < 1$.
      The coordinate system is centred at $(0, 0)$.
      The midpoint of the disc is shifted up to $(0, \delta y)$.
      A (blue) circle is drawn with radius $R_\mathrm{circle}$ (Equation~\ref{eq:minradiusdisc}).
      The circle is then scaled in the $y$ direction by $f_y$, and this scaling is compensated with a scaling in the $x$ direction with $f_x$ (grey).
      The midpoint of the disc is then shifted to $(\delta x, \delta y)$, shearing the ellipse to the proper midpoint (orange).
      From this ellipse the disc parameters can be determined.}
         \label{fig:shallotellipseshearing}
\end{figure}

There are two further points to consider after the disc determination in the previous section.
The first is that five points are required to uniquely determine a given ellipse, so there are an infinite number of ellipses that are centred at $(\delta x,\delta y)$ and intersect with $(\pm0.5,0)$. 
The second is that while the circles defined by $R_\mathrm{circle}$ are the smallest possible \textit{circles} that intersect with the eclipse boundaries, after shearing, the resulting ellipses are not necessarily the ellipse with the smallest semi-major axis, and thus the equations above do not provide the smallest possible discs.
To extend the parameter space defined by the simplest sheared disc model and create a 3 dimensional grid (position of the disc centre $(\delta x, \delta y)$ and the radius scale of the disc, $f_R$), we must be able to model this complete family of ellipses and we do this by introducing two scale factors, $f_x$ and $f_y$.
These scale factors are applied after shifting the midpoint from the origin to $(0,\delta y)$. 
The circle is scaled into an ellipse, by scaling $x$ and $y$ in Equations~\ref{eq:xshear} and~\ref{eq:yshear} as described in Equation~\ref{eq:scaleshear} and visualised in Figure~\ref{fig:shallotellipseshearing}.
\begin{equation}
    \label{eq:scaleshear}
    x \rightarrow x f_x \\
    y \rightarrow y f_y 
\end{equation}
Scaling in one direction, say $y$ using $f_y$, causes the newly formed ellipse to no longer pass through the points $(\pm 0.5,0)$.
For this reason a dependent scaling must be applied in the perpendicular direction, here $x$ with $f_x$, such that the newly formed ellipse once again passes through the points $(\pm 0.5,0)$.
$f_x$ can thus be defined by Equation~\ref{eq:fx} as a function of $f_y$.
\begin{equation}
    \label{eq:fx}
    f_x = f_y \sqrt{\frac{1}{f_y^2 \, \left(1 + 4\, \delta y^2\right) - 4\,\delta y^2}}
\end{equation}
$f_y$ is thus taken as an independent parameter, with $f_x$ compensating the effects of $f_y$ and being a dependent parameter for the ellipse.

Note that the introduction of these scaling factors modifies Equation~\ref{eq:thetamaxmin} into the general case described by Equation~\ref{eq:thetamaxminscale}.
\begin{equation}
    \label{eq:thetamaxminscale}
    \theta = \frac{1}{2} \arctan \left(\frac{2\,f_x\,f_y\,s}{(s^2 + 1)\,f_y^2 - f_x^2 }\right)
\end{equation}
With this it is now possible to analytically determine every ellipse that has a midpoint at $(\delta x, \delta y)$ and passes through the points $(\pm 0.5, 0)$, by adding the scale factor $f_y$.
Thus we have established a three dimensional grid with parameters $\delta x$, $\delta y$ and $f_y$.

\subsection{Converting to Radius Fraction}
\label{subsec:convertingtoradiusfraction}

\begin{figure}
   \centering
   \includegraphics[width=\hsize]{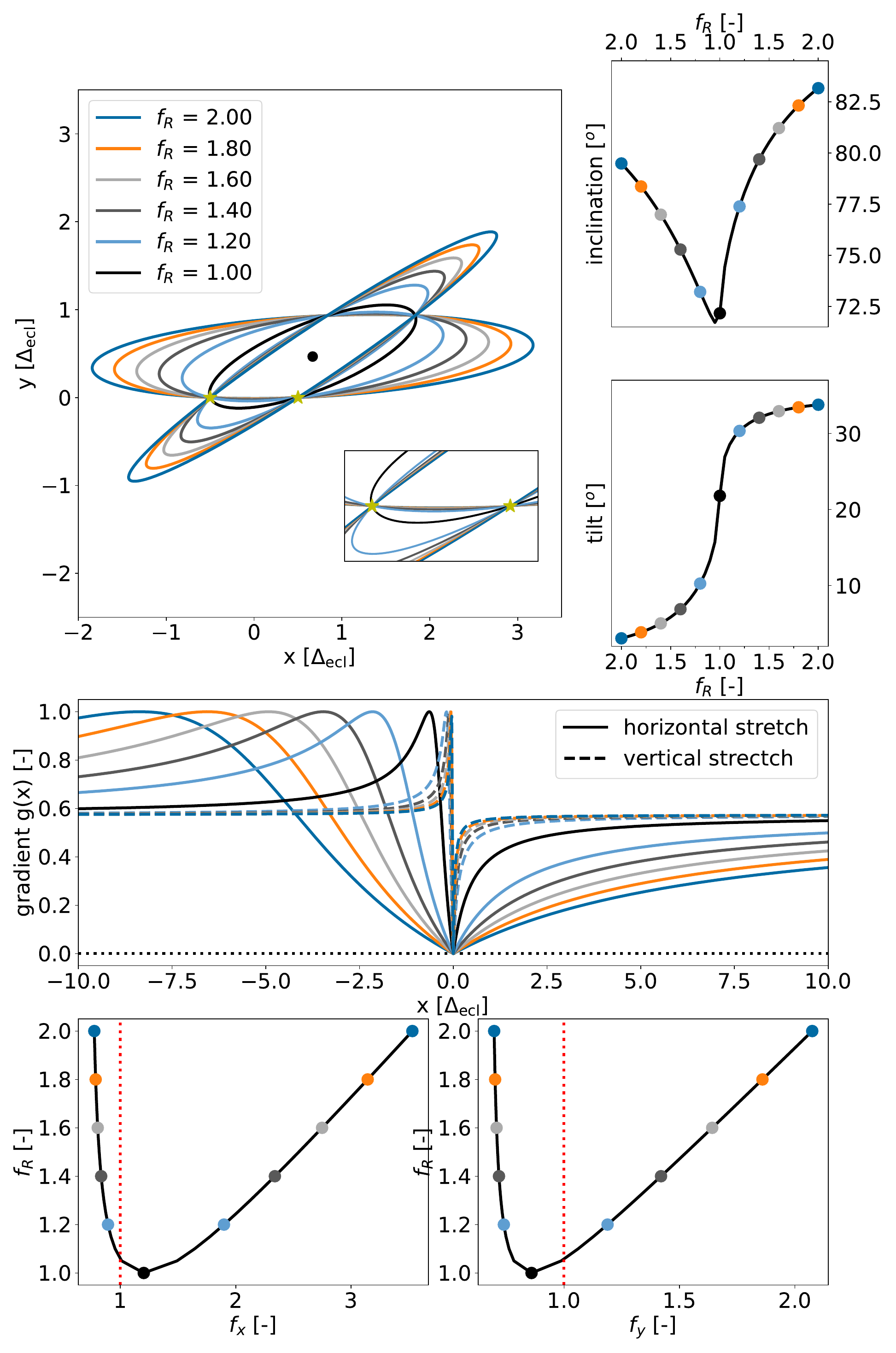}
      \caption{Effects of $f_R$ on Disc Geometry.
      \textit{Top-left:} changes in disc shape and geometry.
      \textit{Top-right:} evolution of inclination with $f_R$.
      \textit{Middle-right:} evolution of tilt with $f_R$.
      \textit{Bottom:} evolution of the theoretical maximum gradient with $f_R$.
      Note that this disc is centred on $(\delta x, \delta y) = (0.67, 0.47)$ in $\Delta_\mathrm{ecl}$.}
         \label{fig:radiusfraction}
\end{figure}

The grid defined above is constructed rapidly as it is determined analytically.
However, $f_y$ is not an intuitive third dimension as it has no obvious physical meaning, so we convert $f_y$ into a the radius scale factor, $f_R = R_\mathrm{disc}/R_\mathrm{min}$, which relates the size of the disc to its minimum radius, at each $(\delta x,\delta y)$.
This dimension drops from any radius scale factor, say $f_{R,\mathrm{max}}$ down to $f_{R,\mathrm{min}} = 1$ back up to $f_{R,\mathrm{max}}$.
This is due to the fact that one can increase the size of the final disc either by stretching the simple circle horizontally, $f_x > f_x \ |_{\, R_\mathrm{min}}$, or vertically, $f_y > f_y \ |_{\, R_\mathrm{min}}$ as shown in Figure~\ref{fig:radiusfraction}.
The analytical formulation of the $R_\mathrm{disc}$ as a function of either $f_x$ or $f_y$ to determine $R_\mathrm{min}$ becomes too involved and is thus determined numerically.
Initially a high resolution grid is generated from the starting $f_x = f_y = 1$ (simplest sheared disc solution) that is extended in one direction with $f_x > 1$ ($f_y < 1$; wider ellipses) and with $f_y > 1$ ($f_x < 1$; taller ellipses) in the other direction.
The $f_R$ grid is determined through linear interpolation of this $f_x$ - $f_y$ grid.

\subsection{Gradient Analysis}
\label{subsubsec:gradientanalysis}

Another source of information is the light curve gradients measured, $\frac{dI(t)}{dt}$.
It should be noted that the light curve of a transiting ring system is not time-symmetric for most geometries (see Figure~\ref{fig:ringsystem}, where the most obvious asymmetry is in the time central occultation between $\sim -0.1 - 0.1$ days).
As described in \citet{Kenworthy_2015}, the light curve gradient is dependent on the size of the star, the local tangent of the ring edge and the direction of motion.
This is characterised by Equation~\ref{eq:gradientkenworthy}, where the projected gradient, $g(t)$, is defined by Equation~\ref{eq:gradient}, and $T_0$ is the transmission at the ingress of the ring edge and $T_1$ is the transmission at the egress of the ring edge.
Note that this equation is only valid when the ring edge boundary can be approximated as a straight line.
\begin{equation}
    \label{eq:gradientkenworthy}
    \frac{dI(t)}{dt} = (T_0 - T_1) \frac{2v_t g(t)}{\pi R_*} \left(\frac{12 - 12 u + 3 \pi u}{12 - 4 u}\right)
\end{equation}
For a given orientation, determined by $\delta x$, $\delta y$, $i$ and $\phi$, an upper bound to the absolute gradient can be calculated by determining what the absolute gradient would be for a transition from fully transparent ($T_0 = 1$) to fully opaque ($T_1 = 0$).
Given this upper bound on the gradient it is possible to rule out a given disc geometry based on the light curve gradients measured.
Those measured gradients must be lower than the upper bounds for the geometry to represent a physically possible solution.

\begin{figure}
   \centering
   \includegraphics[width=\hsize]{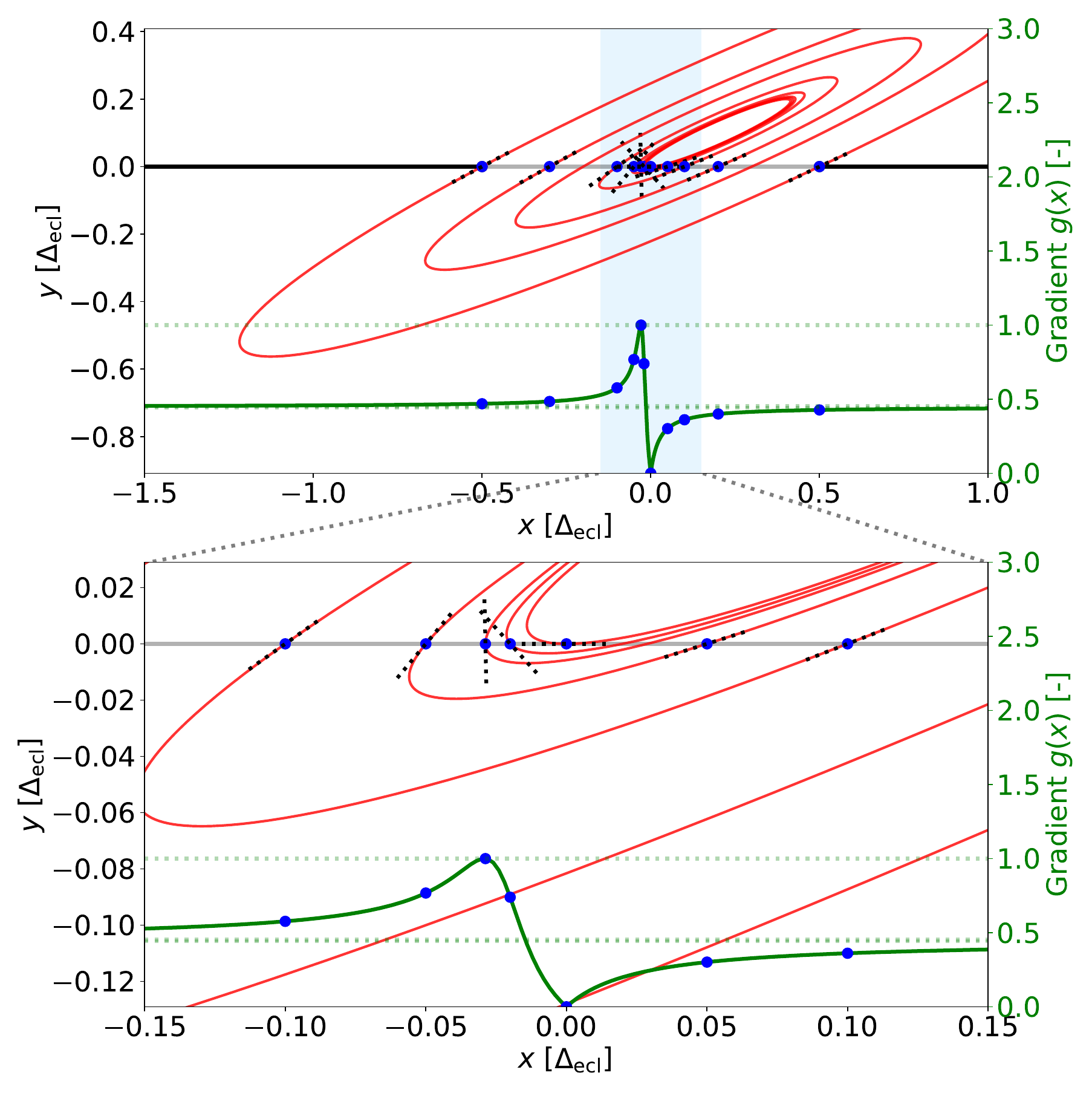}
      \caption{Determining Local Tangents and Projected Gradients.
      \textit{Upper:} shows the full projected gradient curve outside and during the transit. 
      \textit{Lower:} shows the part of the transit where the projected gradient changes significantly and rapidly.
      \textit{Note:} the projected gradient is stable for relatively large $|x|$ because the change in the local ring edge tangent is negligible with position (the left and right asymptote are $1/s$ as dictated by equation~\ref{eq:slope}).
      For small $|x|$ the projected gradient rises to some peak value (where the local tangent is perpendicular to motion), before quickly dropping to 0 (where the local tangent is parallel to motion), and then rising once more.
      In the lower panel it appears that the local tangent rotates counterclockwise with time, which supports the sinusoidal definition of the projected gradient (Equation~\ref{eq:gradient}).}
         \label{fig:gradients}
\end{figure}

We determine the local tangents of the ring edges along the transit path of the star.
To do so, we solve Equation~\ref{eq:scaleshear}, in Cartesian coordinates, for $\frac{dy}{dx} (x)$, producing Equation~\ref{eq:slope}. 
\begin{equation}
    \label{eq:slope}
    \frac{dy''}{dx''} (x) = \frac{f_y^2(s\,{\delta y} + {\delta x} - x)}{s\,f_y^2\,(x - {\delta x}) - {\delta y}\,(s^2\,f_y^2 + f_x^2)}
\end{equation}
This local tangent can be converted to a projected gradient, $g(x)$, by means of Equation~\ref{eq:gradient}. 
This ensures that the projected gradient runs from 1 (perpendicular) to 0 (parallel).
\begin{equation}
    \label{eq:gradient}
    g(x) = \sin\left(\psi (x)\right)\\
    \tan\left(\psi(x)\right) = \frac{dy}{dx} (x)
\end{equation}
A visual representation of the time evolution of the local tangents and their respective projected gradients can be seen in Figure~\ref{fig:gradients}.
The pattern to note is that the projected gradient starts from the $1/s$ asymptote (see equation~\ref{eq:slope}) rising slowly before peaking at some value where the local tangent is perpendicular to the direction of motion ($g(x) = 1$).
There is a steep drop to the time when the local tangent is parallel to the direction of motion ($g(x) = 0$), before the projected gradient rises to the same asymptotic value of $1/s$.

The local tangents of the discs generated by the Shallot Explorer are determined for $(x,\, 0)$. 
$x$ is a position in eclipse durations, which is coupled to the measured light curve gradient, $\frac{dI(t)}{dt}$ measured in $L_* \, \mathrm{day}^{-1}$, time measured in days.
The conversion is described in Equation~\ref{eq:timetoposition}.

\begin{equation}
    \label{eq:timetoposition}
    t = - (x - \delta x)
\end{equation}
Finally we note that the light curve gradients are measured from as many data points that show this trend in the light curve as possible to increase the signal to noise ratio of the measurement. 
This is done by fitting a straight line to the light curve data wherever an extended increase/decrease of flux is observed.

\subsection{Limiting the Parameter Space}
\label{subsec:limitingparameterspace}

The disc models and gradient analysis discussed in the previous sections describe an infinitely large parameter space described by $(\delta x, \, \delta y, \, f_R)$.
To limit the parameter space we introduce two astrophysical restrictions.

\subsubsection{Hill Radius}
\label{subsubsec:hillradius}

An important stability criterion for the disc is that the disc remains stable under the gravitational interactions with the host star.
To ensure an extended lifetime of the disc, $R_\mathrm{disc} < 0.3\,R_\mathrm{Hill}$, where $R_\mathrm{Hill}$ is the Hill radius defined by Equation~\ref{eq:rhill} \citep{Quillen:Trilling:1998}. $a$ is the orbital semi-major axis, $e$ is the eccentricity of the orbit, $m$ is the mass of the companion and $M$ is the mass of the host star.
\begin{equation}
    \label{eq:rhill}
    R_\mathrm{Hill} \approx a\,(1 - e)\,\sqrt[3]{\frac{m}{3M}}
\end{equation}

It is necessary to obtain information ($a, m, M$) that is not readily determined from the normalised light curve.
This also means that the Hill radius restriction applies to the whole grid.

\subsubsection{Gradients}
\label{subsubsec:gradients}

Another consideration is that projected gradients determined from the measured light curve gradients, must be less than the projected gradient upper limits derived from Equations~\ref{eq:gradientkenworthy},~\ref{eq:slope}, and~\ref{eq:gradient}.
This is because the upper limit is based on a boundary from fully opaque ($T = 1$) to fully transparent ($T = 0$), and ring crossings will never produce steeper light curve gradients and thus projected gradients.
As the upper bound of a projected gradient is determined by its location on the $x$ axis for a given geometry (see Figure~\ref{fig:gradients}), the projected gradients measured from the light curve serve as a way to cut out all the unphysical geometries.

\subsubsection{Visualisation}
\label{subsubsec:visualisation}

\begin{figure}
   \centering
   \includegraphics[width=\hsize]{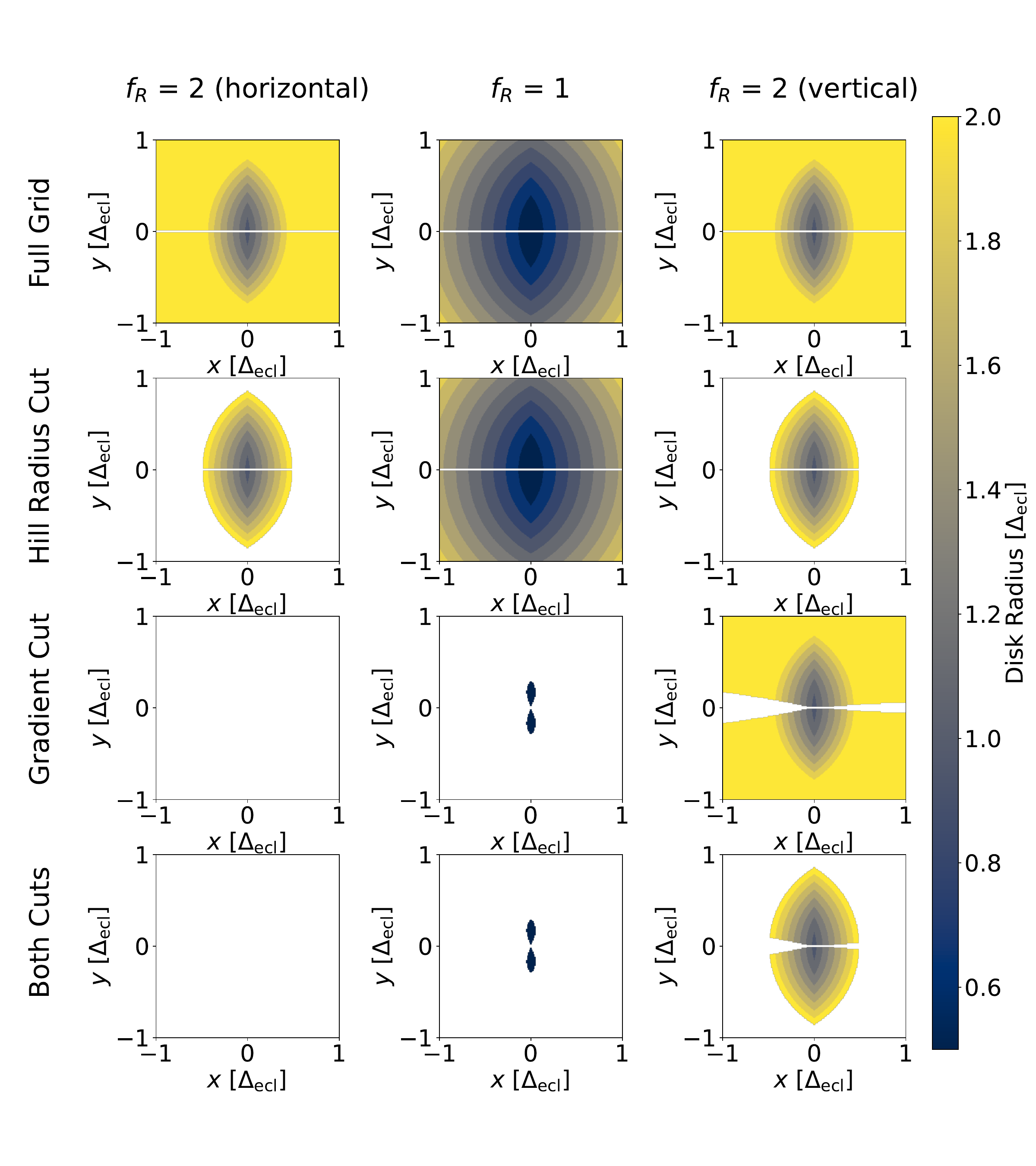}
      \caption{Shallot Visualisation of the Disc Radius. \textit{1$^\mathrm{st}$ row:} shows the full extent of the grid with no restrictions applied. {2$^\mathrm{nd}$ row:} shows the grid after applying a Hill radius cut at $2 \, \Delta_\mathrm{ecl}$. \textit{3$^\mathrm{rd}$ row::} shows the grid after applying a cut due to a projected gradient at $t = 0.45 \, \Delta_\mathrm{ecl}$ with a value of $g(t) = 0.1$. \textit{4$^\mathrm{th}$ row:} shows the grid after applying both the projected gradient and Hill radius cut. \textit{Left:} shows a grid slice with $f_R = 2$ when the disc is stretched horizontally. \textit{Middle:} shows the grid slice with the minimum radius disc ($f_R = 1$). \textit{Right:} shows the grid slice with $f_R = 2$ when the disc is stretched vertically.}
         \label{fig:visualisation}
\end{figure}

Figure~\ref{fig:visualisation} shows the evolution of valid grid points as the restrictions are applied to the grid, specifically the disc radius is shown.
In this case a Hill radius limit of $2 \, \Delta_\mathrm{ecl}$ is applied, and a projected gradient is applied at time $t = 0.45 \, \Delta_\mathrm{ecl}$ with a value $g(t) = 0.1$.
Note that it is from this characteristic visualisation of the disc radii that the Shallot Explorer derives its name.

\subsection{Computational Considerations}
\label{subsec:computations}

To increase the computational efficiency of calculating the grid several degeneracies and symmetries are exploited. 
For a given point $(\delta x, \delta y)$ there exists a point ($\delta_x, -\delta y)$, (-$\delta_x, -\delta y)$ and (-$\delta_x, \delta y)$, which produce ellipses with the same disc radius and inclination, but with a different tilt.
In disc radius and inclination the first quadrant can be reflected off the x and y axis to produce the four quadrants.
Tilt requires an additional transformation after reflecting off the x/y axis.
For quadrant 2 and 4 $\phi \rightarrow 180 - \phi$, and quadrant 1 and 3 remain the same.
This is ensures the tilt is restricted from 0$^\circ$ to 180$^\circ$, which is possible because an ellipse has an axis of symmetry that is aligned with the tilt.
Finally, projected gradients can be reflected off the $x$-axis, but not the $y$-axis, thus requiring the first and second quadrant.

This means that $R_\mathrm{disc}$, $i$ and $\phi$ are calculated for one quadrant, and gradients analysis is performed on two, effectively reducing computation time significantly.

\section{Light Curve Simulations}
\label{sec:simulations}

The transit of ring systems can be modelled to validate whether the geometry of a given grid point of the Shallot Explorer fits the data. These light curves are simulated using the \texttt{pyPplusS} package developed by \citet{Rein:Ofir:2019}.
\texttt{pyPplusS} makes use of a Polygon plus Segments (P+S) algorithm to rapidly and accurately determine the area of a star that is blocked by a ringed planet.
The package fixes the shape of the star to a circle with a radius of 1, which implies that the distances defined are in units of $R_*$.
It is possible to change the intensity profile of the star by applying one of the limb-darkening laws, which can be described with up to four parameters.
The occulter is composed of two objects: the anchor body which is an opaque circle ($T = 0$) with any given size (0 is also possible); and a ring which has an inner radius, outer radius, transmission, tilt and inclination.
It is then possible to model the orbital motion of the occulter by passing in the $(x, \, y)$ coordinate in physical space in units of $R_*$, centred on the midpoint of the star.
The result is that for every given $(x, \, y)$ position, \texttt{pyPplusS} is able to determine the relative intensity of the star (1 when the occulter does not occult the star, 0 when the star is fully blocked by an opaque object).

%
Light curves measure the relative flux of a star w.r.t. time, so the physical space must be converted to temporal space.
This is done by firstly assuming that the path of the occulter across the face of the star (the transit chord) is a straight line, which means we can rotate the coordinate system such that the occulter's $y$ position is fixed, which we denote as the impact parameter (for transits we consider this is justified in Appendix~\ref{sec:derivation}).
The $x$ coordinate now tracks the motion across the transit chord, but this should be converted to time space by introducing a transverse velocity, $v_t$, which is assumed to be constant during the transit of the occulting body and the maximum occultation time, $\delta t$, which is a time-shift parameter to align the light curve.
We are now able to model light curves for planets by setting the transmission of the ring to $T = 1$, discs by setting the planet and inner ring radius to zero, and a ringed planet.

\begin{figure*}
   \resizebox{\hsize}{!}
        {\includegraphics{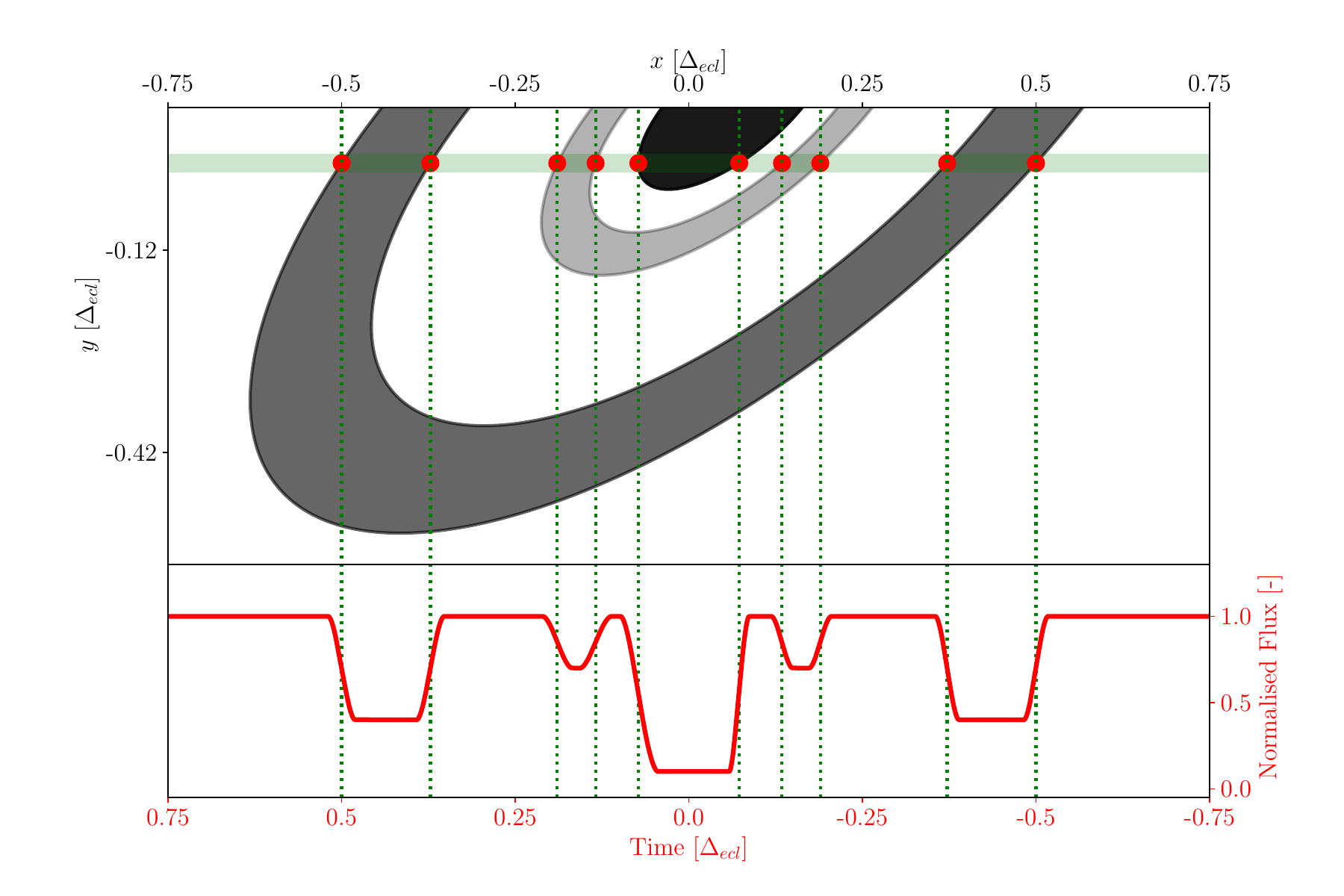}}
  \caption{Geometry of a Transiting Ring System with parameters described in Table~\ref{tab:ringsystemfig} 
  \textit{Top:} shows the ring system in grey scale, with the stellar transit path in green, with a red stellar disc.
  The vertical green dotted lines indicate when the stellar disc is moving behind a ring edge.
  \textit{Bottom:} shows the theoretical light curve for the ring system.
  \textit{Note:} the projected gradient of the light curve is related to the geometrical tangent of the ring edge with the star, and that the light curve drops or rises before the ring crossing because of the finite size of the star.}
     \label{fig:ringsystem}
\end{figure*}

To extend the model to produce simulations of ring system transits, we run the package multiple times.
We initially run it for a planet with no rings and then subsequently run it for each individual ring (with no planet), and then combine the light curves of each component.
Note that this approach fails for small rings that are partially occulted by the anchor body itself, and for transit geometries where the anchor body transits the star and there is a gap between the anchor body and the first ring.
As we are primarily focused on very large ring systems this effect is deemed negligible, especially considering that it is now possible to model ring system transits, and if one were to use enough rings it would even be possible to model a quasi-continuous transmission profile of the disc/ring system.
Figure~\ref{fig:ringsystem} shows an example of a simulation produced by this method.

\begin{table}
    \caption{Parameters for the ring system illustrated in Figure~\ref{fig:ringsystem}}
    \centering
    \begin{tabular}{c c c c}
        \hline\hline\\
        Parameter & Value & Unit\\
        \hline\\
        $R_p$ & 0.003 & $\Delta_\mathrm{ecl}$ & \\
        $\delta x$ & 0.067 & $\Delta_\mathrm{ecl}$ & \\
        $\delta y$ & 0.086 & $\Delta_\mathrm{ecl}$ & \\
        $i$ & 65 & $^\circ$ & \\
        $\phi$ & 40 & $^\circ$ & \\
        $v_t$ & 1.100 & & \\
        $u$ & 0.800 & [-] & \\\\
        \hline\hline\\
        Ring & $R_\mathrm{in} \ [\Delta_\mathrm{ecl}]$  & $R_\mathrm{out} \ [\Delta_\mathrm{ecl}]$ & $T \ [-]$ \\
        \hline\\
        1 & 0.000 & 0.172 & 0.100 \\
        2 & 0.172 & 0.258 & 1.000 \\
        3 & 0.258 & 0.344 & 0.700 \\
        4 & 0.344 & 0.645 & 1.000 \\
        5 & 0.645 & 0.860 & 0.400\\
        \hline
    \end{tabular}
    
    \label{tab:ringsystemfig}
\end{table}

\section{Comparison to Real Data}
\label{sec:validations}

To validate the effective parameter space reduction, the Shallot Explorer is used to determine a viable parameter space for candidate ring systems in the literature.
In both cases we use a relatively coarse Shallot Explorer Grid that extends from $\delta x = -1 \rightarrow 1$, $\delta y = -1 \rightarrow 1$ and $f_R = 5 \, (h) \rightarrow 1 \rightarrow 5 \, (v)$, where $(h)$ is horizontal stretch, and $(v)$ is vertical stretch. 
The bounds of $\delta x$, $\delta y$ and $f_R$ are chosen such that the extremities of the grid at $f_R = 1$ can be cut away by the Hill radius restriction. 
This ensures that the exploration is complete.The grid has a final shape of $(201, 201, 401)$, which is a resolution of $(0.01, 0.01, 0.02)$.
Two examples are explored.

\subsection{J1407b}
\label{subsec:j1407}

\begin{figure}
   \centering
   \includegraphics[width=\hsize]{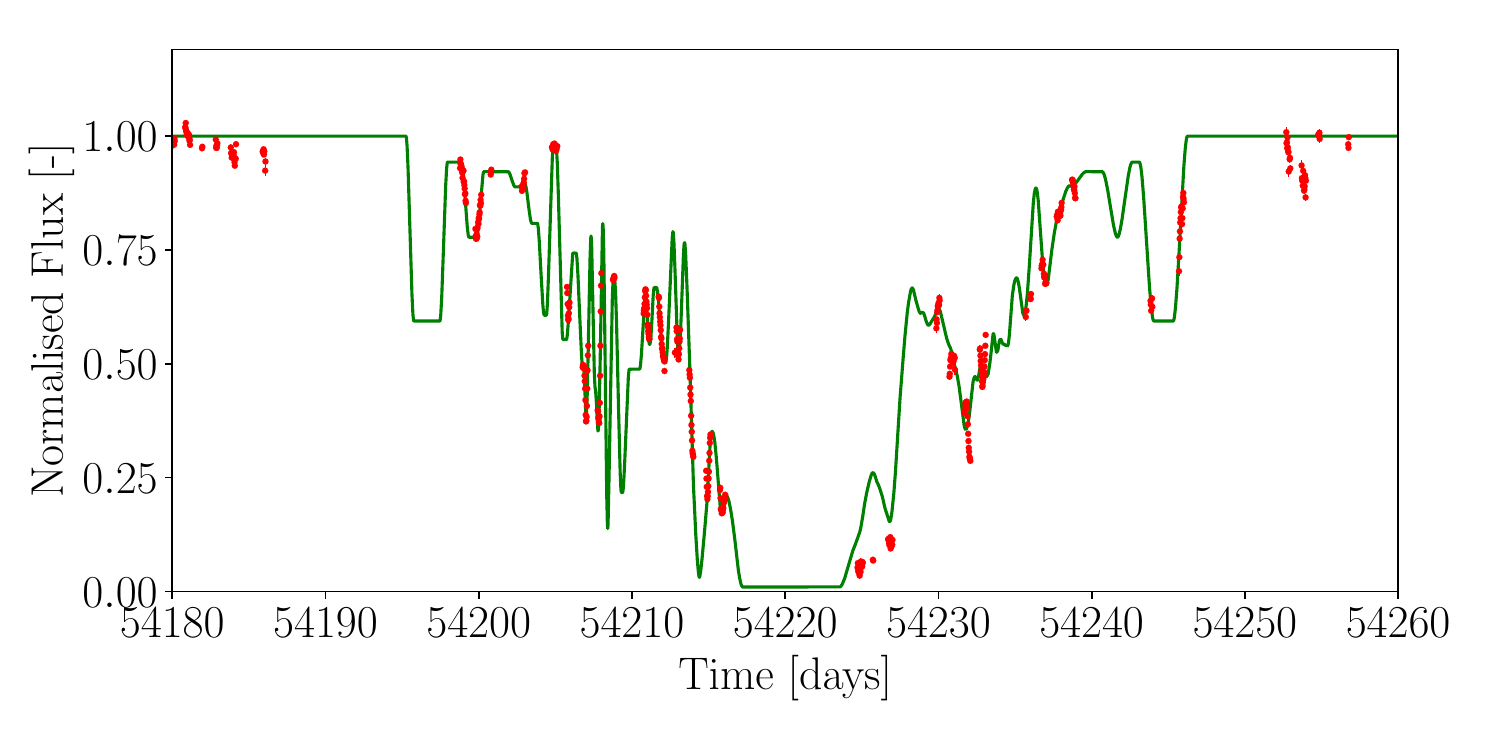}
      \caption{J1407 Light Curve. SuperWASP photometry of the star with the model of the ring system superimposed. The model parameters can be found in Table~\ref{tab:j1407ringparams}. This is a slightly modified version of the top panel of Figure 4 from \citep{Kenworthy_2015}.}
     \label{fig:j1407lightcurve}
\end{figure}

J1407\,b is a widely studied system that highlights the complexity of pinning down and characterising single eclipse systems that exhibit complex sub-structure (see Figure~\ref{fig:j1407lightcurve}).
\begin{table}
    \centering
    \caption{Ring Model Parameters for J1407\,b from \citet{Kenworthy_2015}.}
    \begin{tabular}{c c c c}
        \hline\hline\\
        Parameter & Value & Unit & \\
        \hline\\
        $R_p$ & 0.00 & $\Delta_\mathrm{ecl}$ \\
        $\delta x$ & 0.086 & $\Delta_\mathrm{ecl}$ \\
        $\delta y$ & 0.070 & $\Delta_\mathrm{ecl}$ \\
        $i$ & 70.0 & $^\circ$ \\
        $\phi$ & 166.1 & $^\circ$ \\
        $v_t$ & 262.1 & $R_* \, \Delta_\mathrm{ecl}^{-1}$\\
        $R_\mathrm{Hill}$ & 0.621 & $\Delta_\mathrm{ecl}$\\
        $R_*$ & 0.004 & $\Delta_\mathrm{ecl}$ \\
        $u$ & 0.80 & \textendash \\
        $\Delta_\mathrm{ecl}$ & 50.5 & $\mathrm{day}$\\\\
        \hline\hline\\
        Ring & $R_\mathrm{in} \ [\Delta_\mathrm{ecl}]$  & $R_\mathrm{out} \ [\Delta_\mathrm{ecl}]$ & $T \ [-]$ \\
        \hline\\
        1 & 0.000 & 0.193 & 0.010 \\
        2 & 0.193 & 0.196 & 0.393 \\
        3 & 0.196 & 0.204 & 0.113 \\
        4 & 0.204 & 0.207 & 0.596 \\
        5 & 0.207 & 0.213 & 0.230 \\
        6 & 0.213 & 0.222 & 0.027 \\
        7 & 0.222 & 0.227 & 0.365 \\
        8 & 0.227 & 0.234 & 0.789 \\
        9 & 0.234 & 0.238 & 0.342 \\
        10 & 0.238 & 0.245 & 0.813 \\
        11 & 0.245 & 0.254 & 0.502 \\
        12 & 0.254 & 0.265 & 0.667 \\
        13 & 0.265 & 0.267 & 0.365 \\
        14 & 0.267 & 0.273 & 0.689 \\
        15 & 0.273 & 0.292 & 0.488 \\
        16 & 0.292 & 0.301 & 0.217 \\
        17 & 0.301 & 0.310 & 0.682 \\
        18 & 0.310 & 0.316 & 0.070 \\
        19 & 0.316 & 0.321 & 0.973 \\
        20 & 0.321 & 0.324 & 0.214 \\
        21 & 0.324 & 0.331 & 0.450 \\
        22 & 0.331 & 0.335 & 1.000 \\
        23 & 0.335 & 0.338 & 0.053 \\
        24 & 0.338 & 0.347 & 0.495 \\
        25 & 0.347 & 0.358 & 0.743 \\
        26 & 0.358 & 0.371 & 0.553 \\
        27 & 0.371 & 0.383 & 0.971 \\
        28 & 0.383 & 0.394 & 0.606 \\
        29 & 0.394 & 0.410 & 0.808 \\
        30 & 0.410 & 0.418 & 0.895 \\
        31 & 0.418 & 0.431 & 0.888 \\
        32 & 0.431 & 0.472 & 0.922 \\
        33 & 0.472 & 0.492 & 0.778 \\
        34 & 0.492 & 0.520 & 0.943 \\
        35 & 0.520 & 0.565 & 0.594 \\
        36 & 0.565 & 1.786 & 1.000 \\
        \hline
    \end{tabular}
    \label{tab:j1407ringparams}
\end{table}

\citet{Kenworthy_2015} studied the light curve to fit a ring system and found a 36 ring solution with a reasonable fit. 
Subsequently mass and period limits were studied by \citet{Kenworthy:etal:2015}, followed by constraints on the size and dynamics by \citet{Rieder_2016}.
Due to nature of the eclipse, photographic plates were studied by \citet{Mentel:etal:2018}, and a search for other potential transiting companions was performed by \citet{Barmentloo:etal:2021}.
ALMA data was taken to observe the ring system by \citet{Kenworthy:etal:2020}, which led to the surprising conclusion that J1407\,b is an unbound object that happened to transit J1407 at the right time.
For the purpose of validating the Shallot explorer, we ignore the conclusions reached by \cite{Kenworthy:etal:2020} and direct our attention to the models produced by \citet{Kenworthy_2015}.

To do the gradient analysis we set the eclipse duration to $\Delta_\mathrm{ecl} = 50.5 \ \mathrm{days}$, which is determined from the $t_\mathrm{end} - t_\mathrm{start} = 50.9 \ \mathrm{days}$ of the modelled eclipse (green line from Figure~\ref{fig:j1407lightcurve}) and then subtracting the diameter of the star ($d_* = 0.4 \ \mathrm{days}$). 
The transverse velocity is set to $v_t = 33 \ \mathrm{km}\,\mathrm{s}^{-1}$ and the limb darkening of the star to $u = 0.8$.
To perform the right unit conversions we set the star radius to $R_* = 0.9 \, R_\odot$.
The measured light curve gradients (time $(t)$, light curve gradient $(dI(t)/dt)$, errors, and transmission change $(T_0 - T_1)$) are listed in Table~\ref{tab:j1407gradients}.

\begin{table}
    \caption{J1407 Measured Light Curve Gradients.}
    \label{tab:j1407gradients}
    \centering
    \begin{tabular}{c c c c}
        \hline\hline
         Time & Light Curve & Gradient & Transmission \\
          & Gradient & Error & Change \\
         $[\mathrm{day}]$ & $[L_* \ \mathrm{day}^{-1}]$ & $[L_* \ \mathrm{day}^{-1}]$ & $[-]$ \\
        \hline\\
       -50.645 & -0.160 & 0.020 & -1.00 \\
       -21.745 & -0.160 & 0.020 & -1.00 \\
       -21.565 & -0.410 & 0.030 & -1.00 \\
       -20.665 &  0.270 & 0.009 &  0.25 \\
       -17.745 &  0.160 & 0.030 &  0.15 \\
       -14.835 & -0.970 & 0.060 & -0.40 \\
       -14.735 &  0.620 & 0.070 &  0.40 \\
       -13.695 & -1.200 & 0.040 & -1.00 \\
       -13.555 &  1.700 & 0.040 &  0.70 \\
       -12.835 & -0.320 & 0.040 & -0.70 \\
       -12.685 &  3.000 & 0.060 &  1.00 \\
        -9.835 &  0.520 & 0.050 &  0.70 \\
        -9.615 & -0.520 & 0.010 & -0.70 \\
        -8.825 & -0.540 & 0.030 & -1.00 \\
        -8.625 & -0.270 & 0.010 & -1.00 \\
        -7.685 & -0.570 & 0.030 & -0.50 \\
        -7.545 &  0.660 & 0.060 &  0.50 \\
        -6.765 & -0.870 & 0.020 & -1.00 \\
        -5.765 & -0.860 & 0.070 & -0.80 \\
        -5.605 &  0.870 & 0.020 &  0.80 \\
        -4.875 & -0.880 & 0.100 & -0.85 \\
        -4.685 &  0.170 & 0.010 &  0.85 \\
         9.325 &  0.330 & 0.040 &  1.00 \\
        10.105 &  0.630 & 0.060 &  0.60 \\
        11.105 &  0.320 & 0.050 &  0.50 \\
        11.285 & -0.870 & 0.030 & -0.60 \\
        11.405 & -0.260 & 0.070 & -0.40 \\
        12.125 & -0.580 & 0.030 & -0.60 \\
        12.295 &  0.360 & 0.020 &  0.60 \\
        12.415 &  1.100 & 0.090 &  0.60 \\
        16.255 & -0.150 & 0.010 & -1.00 \\
        18.175 & -0.190 & 0.020 & -1.00 \\
        25.125 &  0.880 & 0.050 &  1.00 \\
        32.205 & -0.280 & 0.030 & -1.00 \\
        33.235 & -0.140 & 0.030 & -1.00 \\
        44.155 & -0.220 & 0.020 & -1.00 \\
        \hline
    \end{tabular}
    \tablefoot{The times, light curve gradients and gradient errors were taken from the \texttt{exorings} repository \footnote{\url{https://github.com/mkenworthy/exorings}} related to \citet{Kenworthy_2015}. The transmission changes were determined as per the \texttt{BeyonCE} repository\footnote{\url{https://github.com/dmvandam/beyonce-shallot}} related to this study.}
\end{table}
\citet{Kenworthy_2015} and \citet{Mentel:etal:2018} show that the ring system exhibited by J1407\,b fills the Hill sphere, thus we take the $R_\mathrm{Hill} = 0.6 \, \mathrm{AU}$, and convert it to time space yielding $R_\mathrm{Hill} = 0.621 \, \Delta_\mathrm{ecl}$. We also perform the gradient analysis twice. Once excluding the measured transmission changes (i.e., setting $(T_0 - T_1) = 1$, which is more inclusive) and a second time including the measured transmission changes.
To illustrate this we rewrite Equation~\ref{eq:gradientkenworthy} in terms of $g(t)$ and introduce the transmission factor, $f_T = (T_0 - T_1)$.
In the rest of this text, transmission scaling refers to the act of using the actual measured transmission factor, $f_T$, instead of substituting unity ($f_T = 1$).

\begin{equation}
    \label{eq:projectedgradienttransmission}
    g(t) = \frac{1}{f_T}\frac{\pi R_*}{2 v_t}\frac{dI (t)}{dt}\left(\frac{12 - 4u}{12 - 12u + 3 \pi u}\right)
\end{equation}
Now it is clear that ignoring the measured transmission ($f_T = 1$) provides the lowest possible $g(t)$, which means the fewest ring system configurations are excluded.
Note also that if, for any reason (e.g. sparse photometry), it is not possible to measure $f_T$ then the value defaults to 1. 
This provides an inclusive analysis, and by applying the measured $f_T$ the analysis becomes more exclusive.

\begin{figure}
   \centering
   \includegraphics[width=\hsize]{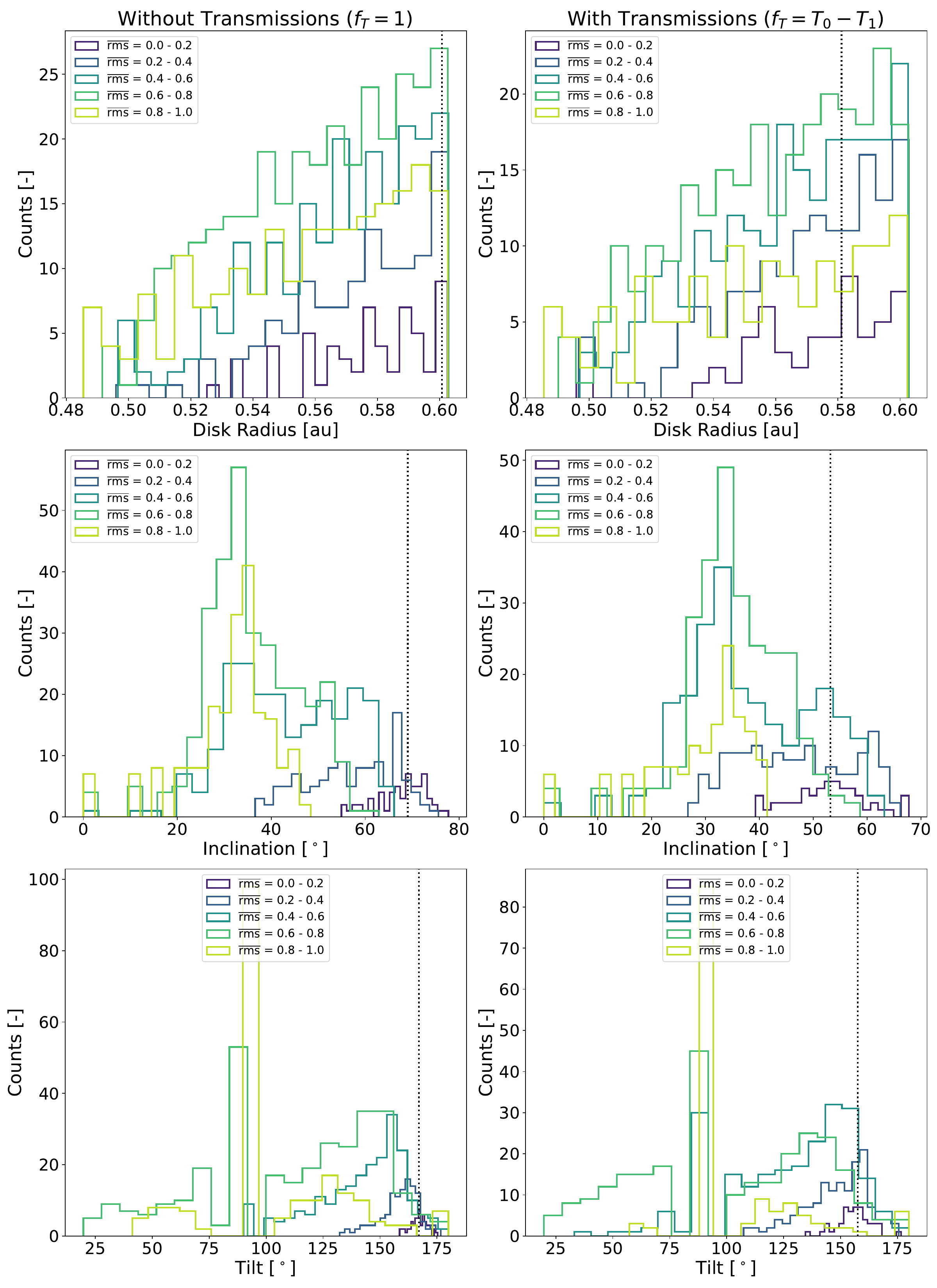}
      \caption{J1407 Disc Property Distributions. Each panel shows the distribution of that particular disc property that has been grouped by $\overline{\mathrm{rms}}$ ($\overline{\mathrm{rms}}$ runs from 0 - 1) bins. The vertical line shows the solution with the smallest $\overline{\mathrm{rms}}$ \textit{Top:} Disc Radius, \textit{Middle:} Inclination and \textit{Bottom:} Tilt. Note that the tilt distribution is not bimodal because we only show solutions with a positive impact parameter.}
     \label{fig:j1407features}
\end{figure}

Once the gradient analysis and Hill sphere restriction have been performed we use the $\chi^2$ goodness of fit as described in Equation~\ref{eq:gradientfit}, where $o$ is the observed (measured) value and $e$ is the expected value and $\sigma$ is the error on the measurement, to rank the physically possible configurations.

\begin{equation}
    \label{eq:gradientfit}
    \chi^2 = \displaystyle\sum_{i=1}^{n} \frac{(g_{i, \, o} - g_{i, \, e})^2}{\sigma_i^2}
\end{equation}
The $\chi^2$ value is determined for each grid point (note that all points where the measured value is greater than the expected value are ignored as they are physically impossible).
These points are then extracted and sorted by the $\chi^2$ value.
We choose to investigate all solutions, and find that from the 16,200,801 explored grid points, only 1,850 are physically possible.
These solutions are are then sorted by disc radius.
Solutions with $\delta y < 0$ are ignored as they are degenerate (see Section~\ref{subsec:computations}).
To visualise the distribution of these physical solutions, we introduce the normalised r.m.s., $\overline{\mathrm{rms}}$, described in Equation~\ref{eq:rms}, where $\chi^2_\mathrm{min}$ and $\chi^2_\mathrm{max}$ are the smallest and largest $\chi^2$ values measured respectively.

\begin{equation}
    \label{eq:rms}
    \overline{\mathrm{rms}} = \frac{\chi^2 - \chi^2_\mathrm{min}}{\chi^2_\mathrm{max} - \chi^2_\mathrm{min}}
\end{equation}
The solutions are subsequently placed into 5 bins with a width of 0.2, running from $\overline{\mathrm{rms}} = 0.0 - 1.0 $ and the distribution of disc parameters is presented in Figure~\ref{fig:j1407features}.

Figure~\ref{fig:j1407features} demonstrates the complexity of the possible physical systems that could fit the light curve data.

The reason why we perform the gradient analysis while ignoring the transmission factor is due to the large uncertainties involved in determining them.
This is primarily due to the photometric data being too sparse to confidently measure a transmission change.
To accurately determine $f_T$ for a particular light curve gradient one must have a flat-bottom before and after the gradient time, which in most cases is not available.
This leads to the estimation of $f_T$. One must take the largest upper bound of $f_T$, because underestimating it can lead to unphysical projected gradients ($g(t) > 1$).

\begin{figure}
    \centering
    \includegraphics[width=\hsize]{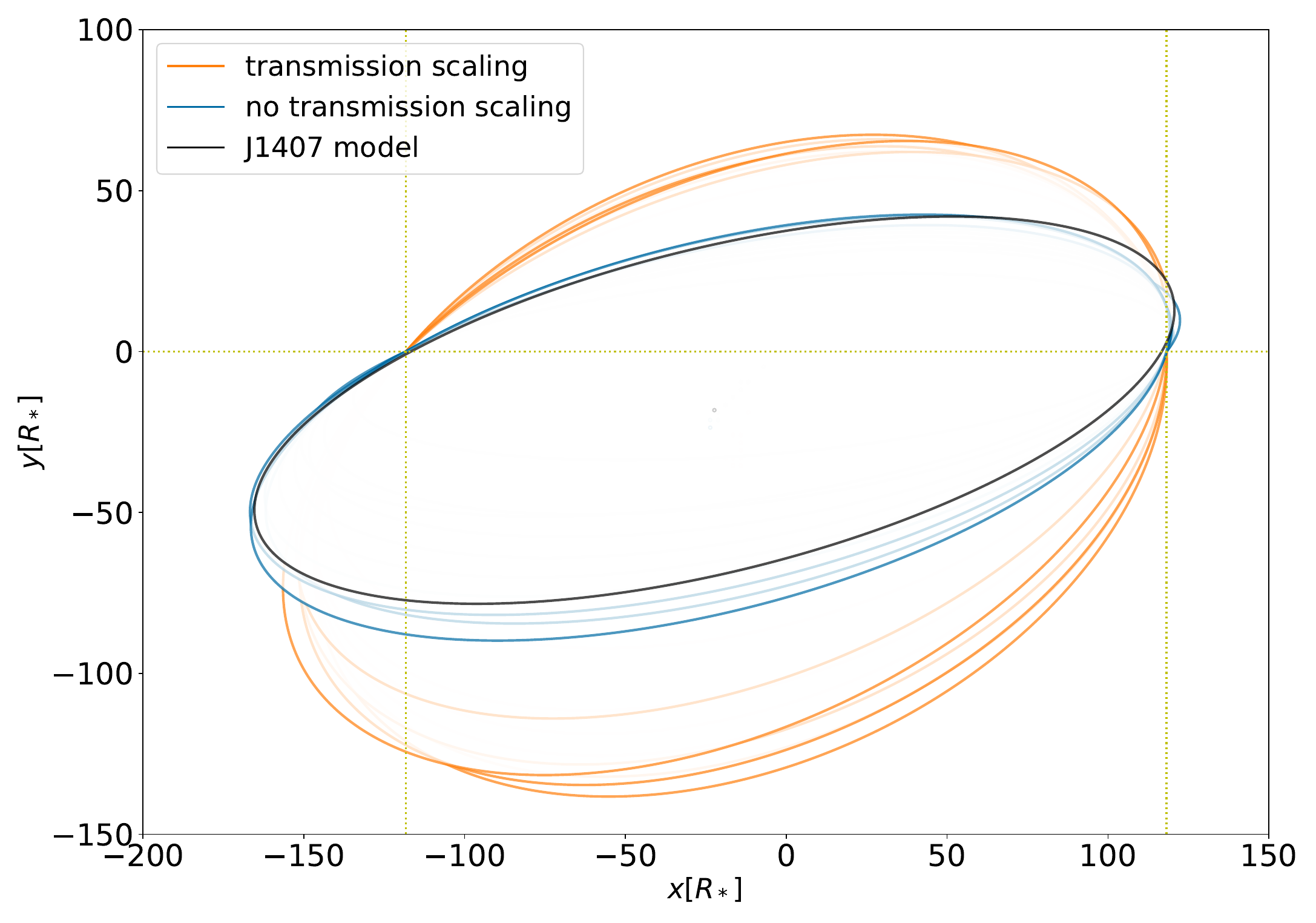}
    \caption{J1407 Disc Models. The results of the Shallot Explorer are shown along with the original model for the J1407 disc (black). In blue, solutions that ignore transmission scaling, $f_{T,x} = 1$, and in orange the solutions that do.}
    \label{fig:j1407ringsystems}
\end{figure}

\begin{figure}
   \centering
   \includegraphics[width=\hsize]{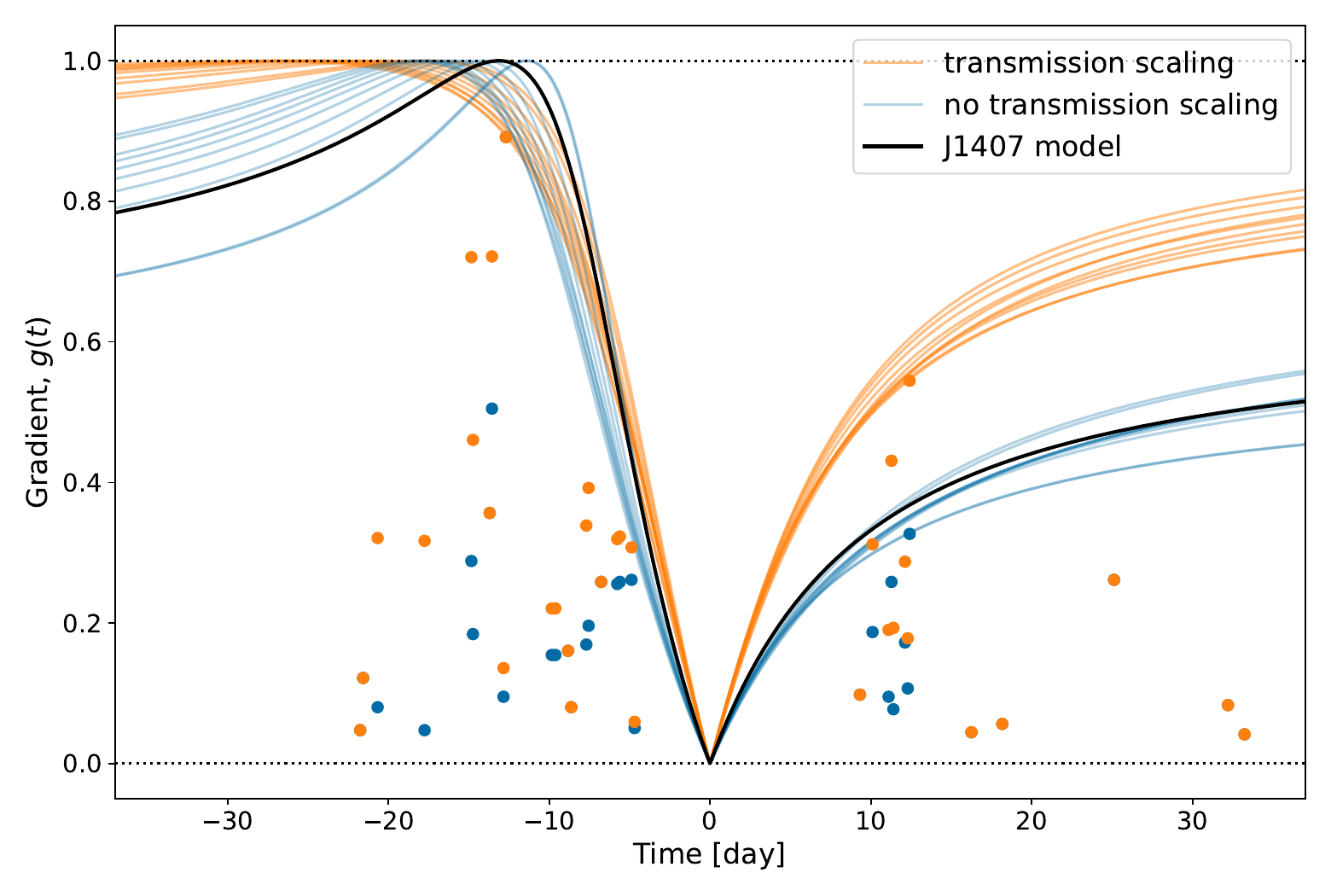}
      \caption{J1407 Projected Gradients. The projected gradients are plotted for the ring system models depicted in Figure~\ref{fig:j1407ringsystems}. Note that the orange data points make use of $f_T$ and that the J1407 model (black) does not ($f_T = 1$).}
     \label{fig:j1407gradients}
\end{figure}

The ring systems of both analyses are visible in Figure~\ref{fig:j1407ringsystems} with projected gradient information shown in Figure~\ref{fig:j1407gradients}. 
Notice how there are some solutions which have a smaller disc radius than the original paper.
This highlights the purpose of Shallot Explorer in finding physically smaller ring solutions.

Table~\ref{tab:modelcomp} compares the smallest disc solution found by Shallot Explorer with a $\overline{\mathrm{rms}} < 0.2$ to the solution determined by \citet{Kenworthy_2015}.

\subsection{PDS 110}
\label{subsec:pds110}

PDS 110 (HD 290380) is a young ($\sim11$ Myr old) T-Tauri star in the Orion OB1 Association.
It exhibited two extended, deep eclipses with a duration of approximately 25 days, which were separated by about 808 days.
\begin{figure}
    \centering
    \includegraphics[width=\hsize]{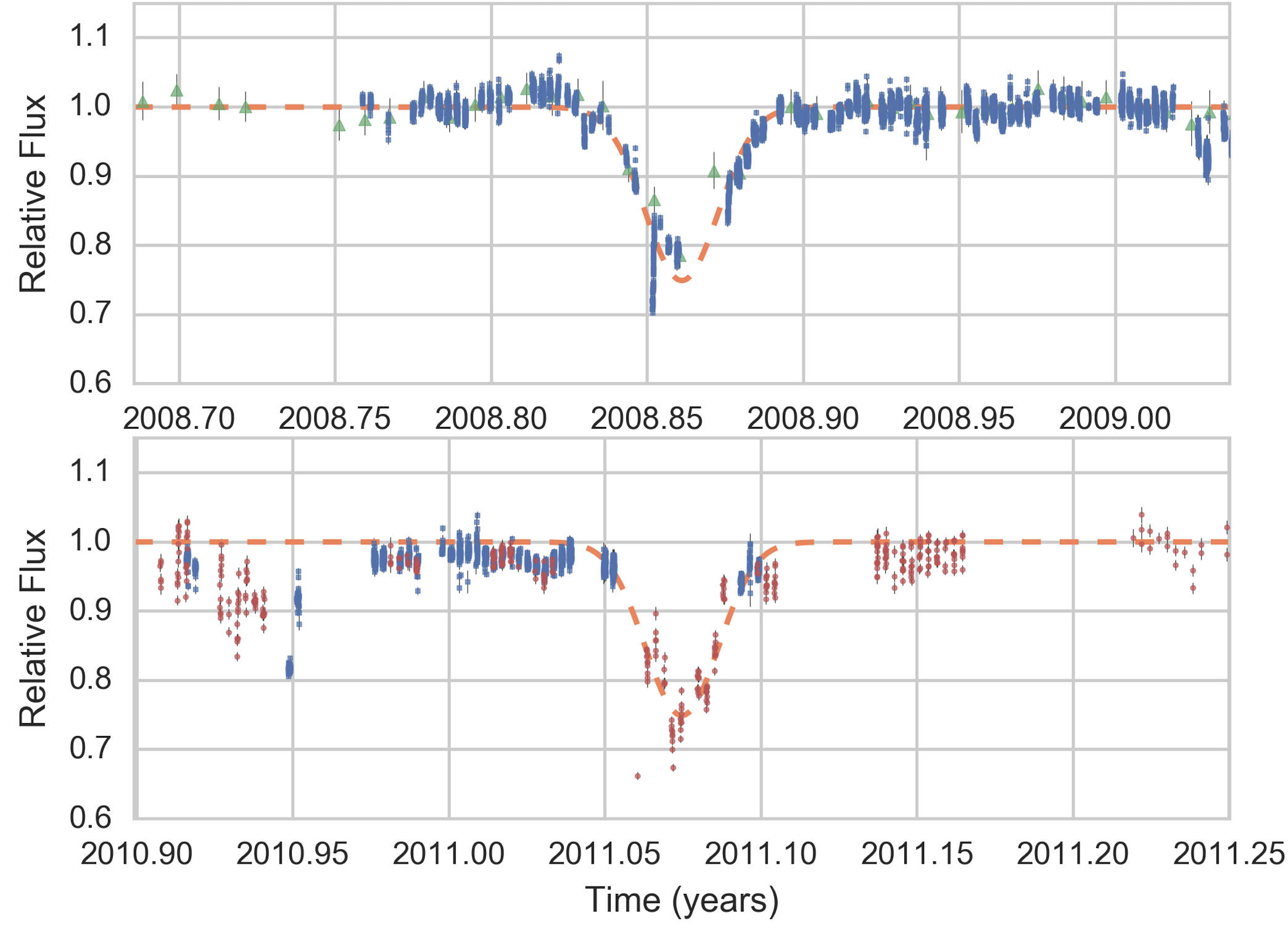}
    \caption{PDS 110 Light Curve. \textit{Top}: shows the first eclipse from 2008. \textit{Bottom}: shows the second eclipse from 2011. The red circles are data from KELT, the blue squares is data from SuperWASP, the green triangles are data from ASAS. The best fit eclipse model is overplotted in orange with parameters in Table~\ref{tab:pds110ringparams}. This figure is extracted from the bottom left panels of Figure 1 from \citep{Osborn2017}.}
    \label{fig:pds110lightcurve}
\end{figure}
\begin{table}
    \centering
    \caption{Ring Model Parameters for PDS110 from \citet{Osborn2017}.}
    \begin{tabular}{c c c c}
        \hline\hline\\
        Parameter & Value & Unit & \\
        \hline\\
        $R_p$ & 0.00 & $\Delta_\mathrm{ecl}$ \\
        $\delta x$ & 0.161 & $\Delta_\mathrm{ecl}$ \\
        $\delta y$ & 0.098 & $\Delta_\mathrm{ecl}$ \\
        $i$ & 74.7 & $^\circ$ \\
        $\phi$ & 158.6 & $^\circ$ \\
        $v_t$ & 39.89 & $R_* \, \Delta_\mathrm{ecl}^{-1}$\\
        $R_\mathrm{Hill}$ & 1.741 & $\Delta_\mathrm{ecl}$\\
        $R_*$ & 0.025 & $\Delta_\mathrm{ecl}$ \\
        $u$ & 0.80 & \textendash \\
        $\Delta_\mathrm{ecl}$ & 25 & $\mathrm{day}$\\\\
        \hline\hline\\
        Ring & $R_\mathrm{in} \ [\Delta_\mathrm{ecl}]$  & $R_\mathrm{out} \ [\Delta_\mathrm{ecl}]$ & $T \ [-]$ \\
        \hline\\
         1 & 0.000 & 0.160 & 0.750 \\
         2 & 0.160 & 0.197 & 0.768 \\
         3 & 0.197 & 0.234 & 0.920 \\
         4 & 0.234 & 0.271 & 0.550 \\
         5 & 0.271 & 0.309 & 1.000 \\
         6 & 0.309 & 0.346 & 0.500 \\
         7 & 0.346 & 0.395 & 0.700 \\
         8 & 0.395 & 0.420 & 0.875 \\
         9 & 0.420 & 0.457 & 0.900 \\
        10 & 0.457 & 0.494 & 1.000 \\
        11 & 0.494 & 0.531 & 0.700 \\
        12 & 0.531 & 0.697 & 0.970 \\
        13 & 0.697 & 0.758 & 0.860 \\
        14 & 0.758 & 0.803 & 0.982 \\
        15 & 0.803 & 0.840 & 1.000 \\
        \hline
    \end{tabular}
    \label{tab:pds110ringparams}
\end{table}
These eclipses were studied by \citet{Osborn2017} and were interpreted as the result of the transit of a circumsecondary disc around an unseen companion, PDS 110\,b.
An observing campaign was subsequently setup to detect an expected transit event, but no such event occurred \citep{Osborn:etal:2019}.
We focus on the ring models produced by \citet{Osborn2017}.

The analysis follows the same steps as with J1407 in the previous section, but we limit the number of solutions investigated to 10,000 because as the number of solutions increases, they become so general that they become meaningless.
The Hill radius for PDS 110 is 0.69 $\mathrm{AU}$ and to do the gradient analysis the eclipse duration is set to $\Delta_\mathrm{ecl} = 25 \ \mathrm{days}$, the transverse velocity is set to $v_t = 27 \ \mathrm{km}\,\mathrm{s}^{-1}$ and the limb darkening of the star to $u = 0.8$.
To perform the right unit conversions we set the star radius to $R_* = 2.23 \, R_\odot$.
The measured light curve gradients (time $(t)$, light curve gradient $(dI(t)/dt)$, errors and transmission change $(T_0 - T_1)$) are listed in Table~\ref{tab:pds110gradients}.
\begin{table}
    \caption{PDS 110 Measured Light Curve Gradients.}
    \label{tab:pds110gradients}
    \centering
    \begin{tabular}{c c c c}
        \hline\hline
         Time & Light Curve& Gradient & Transmission \\
          & Gradients & Error & Change \\
         $[\mathrm{day}]$ & $[L_* \ \mathrm{day}^{-1}]$ & $[L_* \ \mathrm{day}^{-1}]$ & $[-]$ \\
        \hline
        -14.20 &  0.24 & 0.03 &  1.00 \\
        -13.10 & -0.03 & 0.05 & -1.00 \\
        -12.18 &  0.05 & 0.02 &  1.00 \\
        -11.13 & -0.36 & 0.02 & -1.00 \\
        -10.17 &  0.03 & 0.01 &  0.97 \\
         -9.97 &  0.05 & 0.01 &  0.97 \\
         -8.97 &  0.04 & 0.03 &  0.97 \\
         -5.18 & -0.02 & 0.02 & -0.90 \\
         -5.00 & -0.37 & 0.07 & -1.00 \\
         -4.00 &  0.19 & 0.35 &  0.90 \\
         -3.06 &  0.65 & 0.03 &  1.00 \\
         -2.98 &  0.11 & 0.44 &  0.85 \\
         -2.05 & -0.54 & 0.08 & -1.00 \\
         -1.27 & -0.02 & 0.05 & -0.80 \\
         -1.04 &  0.23 & 0.09 &  1.00 \\
         -0.21 &  0.04 & 0.01 &  0.80 \\
          0.96 &  0.03 & 0.09 &  0.80 \\
          1.95 & -0.04 & 0.07 & -0.80 \\
          2.94 &  0.07 & 0.13 &  0.88 \\
          3.95 & -0.01 & 0.08 & -0.95 \\
          5.77 &  0.24 & 0.02 &  1.00 \\
          6.00 &  0.03 & 0.01 &  0.95 \\
          6.85 & -0.02 & 0.00 & -0.90 \\
          7.91 & -0.01 & 0.01 & -0.95 \\
          7.97 &  0.01 & 0.02 &  0.97 \\
          8.84 & -0.01 & 0.01 & -0.95 \\
          8.95 &  0.12 & 0.09 &  0.97 \\
          9.78 &  0.04 & 0.01 &  0.95 \\
          9.95 &  0.06 & 0.14 &  1.00 \\
         11.78 &  0.03 & 0.01 &  1.00 \\
         13.83 & -0.05 & 0.01 & -1.00 \\
         14.77 &  0.02 & 0.01 &  1.00 \\
         15.78 &  0.00 & 0.01 &  1.00 \\

        \hline
    \end{tabular}
    \tablefoot{The times, light curve gradients and gradient errors were taken from the \texttt{pds\_110\_exorings} repository\footnote{\url{https://github.com/mkenworthy/pds\_110\_exorings}} related to \citet{Osborn2017}.}
\end{table}

\begin{figure}
    \centering
    \includegraphics[width=\hsize]{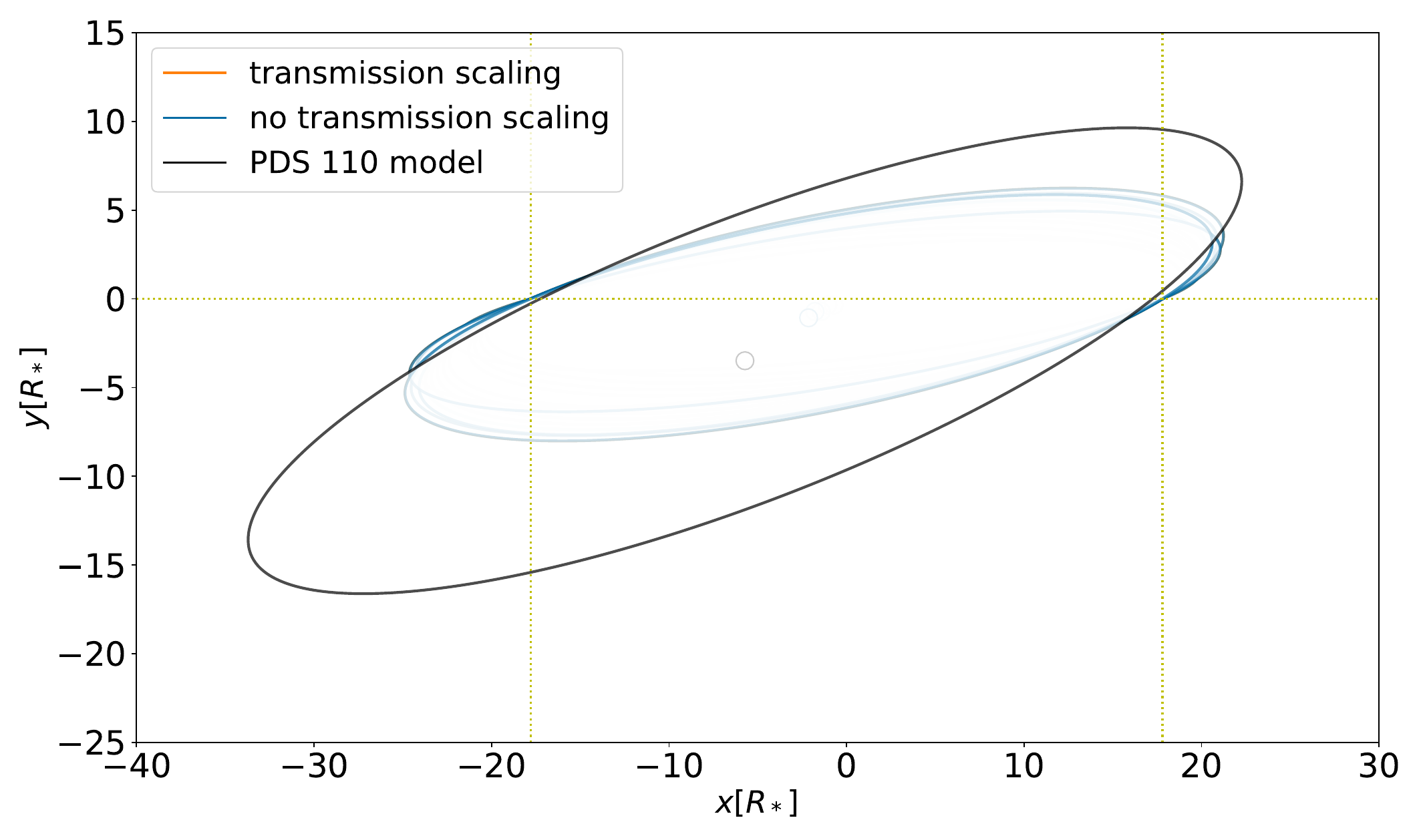}
    \caption{PDS 110 Disc Models.
    The results of the Shallot Explorer are shown along with the original model for the PDS 110 disc (black).
    In blue, solutions that ignore transmission scaling, $(T_0 - T_1)_x = 1$, and in orange the solutions that do.}
    \label{fig:pds110ringsystems}
\end{figure}

\begin{figure}
   \centering
   \includegraphics[width=\hsize]{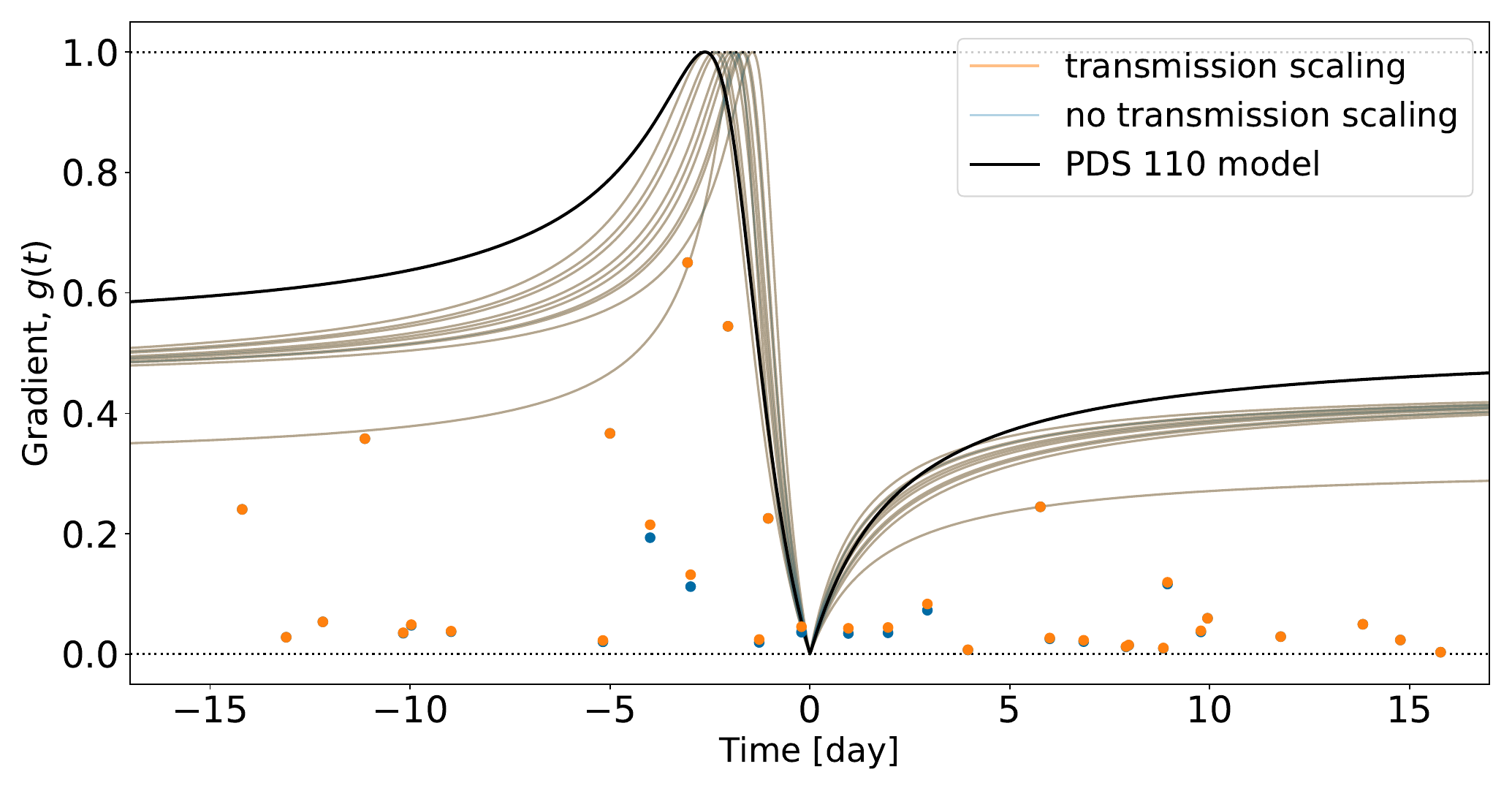}
      \caption{PDS 110 Projected Gradients.
      The projected gradients are plotted for the ring system models depicted in Figure~\ref{fig:j1407ringsystems}.
      Note that the orange data points make use of $f_T$ and that the PDS110 model (black) does not ($f_T = 1$).}
     \label{fig:pds110gradients}
\end{figure}

The ring systems of both passes are visible in Figure~\ref{fig:pds110ringsystems} with projected gradient information shown in Figure~\ref{fig:pds110gradients}. 
Notice how all the solutions have a significantly smaller disc radius than the original model used.
This demonstrates the use case of the Shallot Explorer.

\begin{figure}
   \centering
   \includegraphics[width=\hsize]{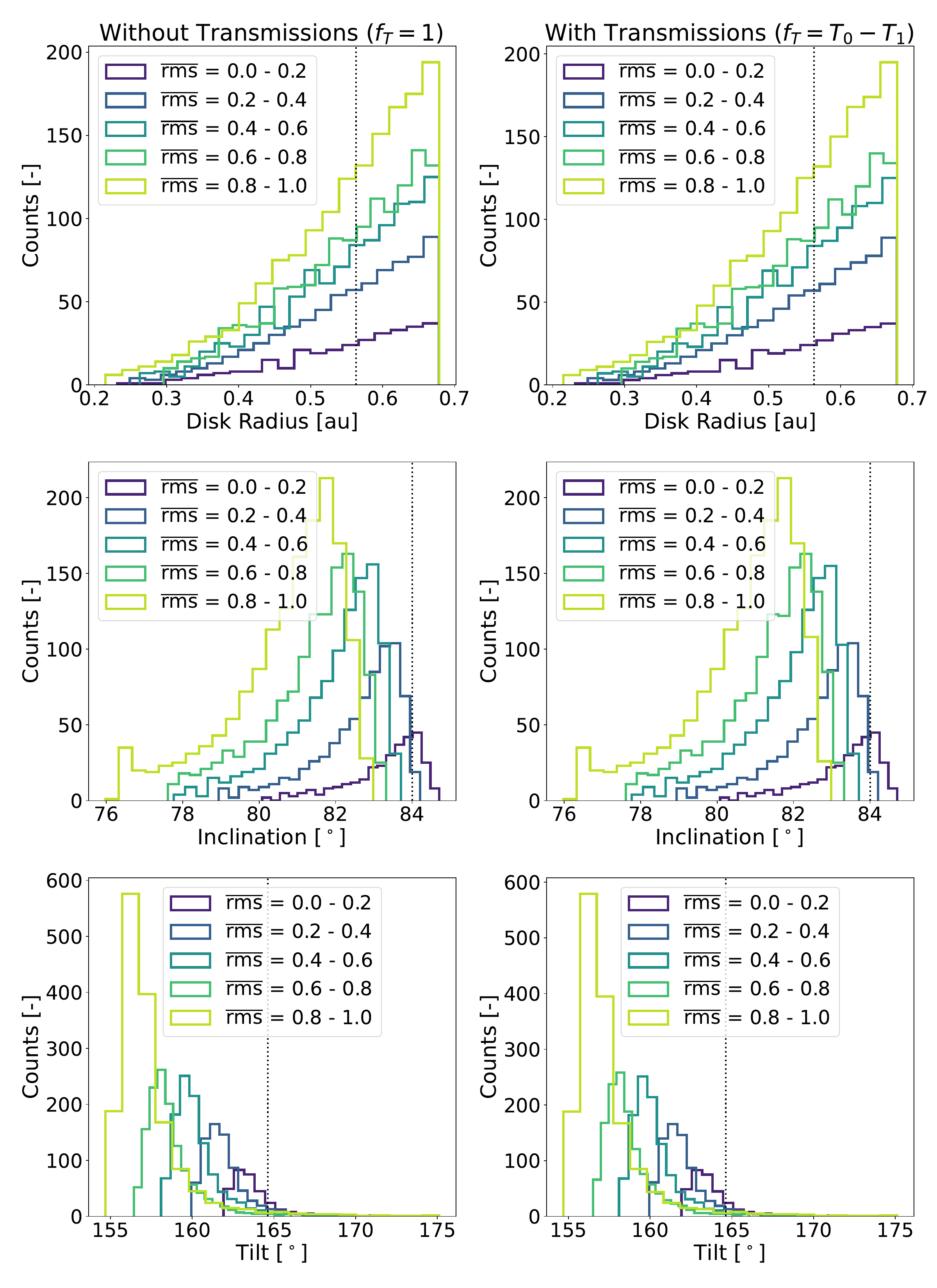}
      \caption{PDS 110 Disc Property Distributions. Each panel shows the distribution of that particular disc property that has been grouped by $\overline{\mathrm{rms}}$ ($\overline{\mathrm{rms}}$ runs from 0 - 1) bins. The vertical line shows the solution with the smallest $\overline{\mathrm{rms}}$ \textit{Top:} Disc Radius, \textit{Middle:} Inclination and \textit{Bottom:} Tilt. Note that the tilt distribution is not bimodal because we only show solutions with a positive impact parameter.}
     \label{fig:pds110features}
\end{figure}

As with J1407, we also show the distribution of the parameters for bins of $\overline{\mathrm{rms}}$ in Figure~\ref{fig:pds110features}, and we see in Table~\ref{tab:modelcomp} how the smallest disc with an $\overline{\mathrm{rms}} < 0.2$, compares with the model determined by \citet{Osborn2017}.

\begin{table}
    \centering
    \renewcommand{\arraystretch}{1.5}
    \caption{Literature vs. Shallot Explorer Model Comparisons. The value listed under Shallot is the smallest disc solution with a $\overline{\mathrm{rms}} < 0.2$ }
    \begin{tabular}{l | c c}
        \hline\hline
        \textbf{J1407\,b} & \citet{Kenworthy_2015} & Shallot \\
        \hline
        $R_\mathrm{disc} \ \left[ \mathrm{AU} \right]$ & 0.60 & 0.52 \\
        $i \ \left[^\circ\right]$ & 70 & 78 \\
        $\phi \ \left[^\circ\right]$ & 166 & 175 \\
        $\delta x \ \left[ \mathrm{AU} \right]$ & 0.092 & 0.028 \\
        $\delta y \ \left[ \mathrm{AU} \right]$ & 0.075 & 0.019 \\

        \hline
        \hline
        \textbf{PDS 110} & \citet{Osborn2017} & Shallot \\
        \hline
        $R_\mathrm{disc} \ \left[ \mathrm{AU} \right]$ & 0.67 & 0.22 \\
        $i \ \left[^\circ\right]$ & 75 & 82 \\
        $\phi \ \left[^\circ\right]$ & 159 & 175 \\
        $\delta x \ \left[ \mathrm{AU} \right]$ & 0.063 & 0.028 \\
        $\delta y \ \left[ \mathrm{AU} \right]$ & 0.038 & 0.019 \\
        \hline
    \end{tabular}
    \label{tab:modelcomp}
\end{table}

\section{Discussion and Conclusions}
\label{sec:discussionandconclusions}

The Shallot Explorer module of the \texttt{BeyonCE} package is an effective tool at exploring and visualising the parameter space spanned by circumsecondary discs.
It produces a grid that is described with the offset of the centre of the disc w.r.t. to the centre of the eclipse in time space with a final dimension describing the size of the disc relative to the minimum possible size at that specific centre offset.
The parameter space is nevertheless very large, but can be limited by two fundamental properties.
The first is the Hill radius, which imposes a maximum size of the disc based on stability criteria for the object orbiting the host star.
The second is the relationship between the light curve gradients measured and the physical local tangent of the edge of the disc crossing the star.
Implementing these two cuts effectively, significantly reduces the parameter space to much more manageable starting points for further analysis.
Importantly, the Shallot Explorer makes no use of prior distributions.
Instead it is designed to explore the unbiased parameter space of everything that is physically possible.
We are producing a systematic approach and defining the boundary between possible and not possible disc geometries based on the gradients as opposed to only minimising the radius of the disc. 

Two examples were explored.
The first is J1407\,b, an object with a very large ring system proposed.
Using the information obtained from the light curve (light curve gradients, eclipse duration, ... etc.) and from the orbits of the system (mass of the star, orbital mechanics), Shallot Explorer was capable of returning solutions of a similar geometry as to that found by \citet{Kenworthy_2015} (see Table~\ref{tab:modelcomp}).
The same analysis was done for the PDS 110 system, producing solutions
that are significantly smaller than the proposed disc (see Table 7
and Figure 13). 
This analysis generally provides a robust exploration of the disc orientations and in these particular cases (J1407b and PDS 110) reduces the parameter space to $\sim$ 10\% of our initial grid.
Figure~\ref{fig:j1407features} and~\ref{fig:pds110features} show the distribution of each disc property for the valid configurations with a positive impact parameter, for J1407 and PDS 110, respectively.
By valid configurations we refer to the solutions left over after applying the Hill radius and projected gradient cuts.
We choose the positive impact parameter because we obtain the same disc solution by flipping the sign of the impact parameter and reflecting the tilt off the $\phi = 90^\circ$ axis (as described in Section~\ref{sec:parameters} and~\ref{subsec:computations}). 

Both of these say something different about Shallot Explorer.
The fact that J1407\,b has such similar solutions implies that Shallot Explorer is able to find a solution that is determined by other means and that the solution found for the J1407 system is one of the smallest solutions given the light curve data.
The fact that PDS 110 shows smaller solutions implies that Shallot Explorer is able to find solutions with a better fit than a focused case study.
The more Shallot Explorer is used on similar complex systems, the more evidence will amount to its use as the starting point in modelling of complex light curves.

While the grid produced is linearly scalable, the implications do have hidden consequences.
Consider the light curve for J1407\,b (Figure~\ref{fig:j1407lightcurve}): it shows no definitive eclipse duration as we are missing a clear ingress and egress and confirmation that the light curve remains steady outside of this time period.
While not knowing the eclipse duration has no effect on the validity of the grid points, it does have an influence on the gradient cuts applied.
This becomes obvious when looking at Equation~\ref{eq:timetoposition}. 
The time measured changes w.r.t. eclipse duration, which changes the maximum theoretical value that gradient could have.
For gradients measured close to the ingress and egress this will have a small effect.
For gradients measured closer to the times when the local tangent of the ring edge is parallel and perpendicular to the direction of motion (between $-0.1$ and $0.1$ in Figure~\ref{fig:gradients}), the uncertainty in eclipse duration could have a very significant effect.

Another limitation that is not considered here is that every ring crossing produces four gradients: two sets of an ingress and an egress.
This is not considered when making cuts to the parameter space, which ensures that the parameter space is larger than it could be.
Shallot Explorer is thus inclusive of the correct solutions; however, determining the best fit solutions as described by Equation~\ref{eq:gradientfit} is only a starting point when moving on to ring system, light curve fitting. 

It is also useful to investigate the $R_\mathrm{disc}$ value at which the Shallot Explorer fails to produce viable results due to the breakdown of the first assumption $$h_\mathrm{disc} << R_* << R_\mathrm{disc}$$

\texttt{BeyonCE} will be supplemented with a second module that will be used for the fitting of the actual ring system transmissions itself, based on results exported from the Shallot Explorer or unique manual input from the user.

\begin{acknowledgements}
      To achieve the scientific results presented in this article we made use of the \emph{Python} programming language\footnote{Python Software Foundation, \url{https://www.python.org/}}, especially the \emph{SciPy} \citep{virtanen2020}, \emph{NumPy} \citep{numpy} and \emph{Matplotlib} \citep{Matplotlib}.
\end{acknowledgements}

\begin{appendix}

\section{Validity of {\tt exoring} software geometry}
\label{sec:derivation}

\begin{figure}
   \centering
   \includegraphics[width=\hsize]{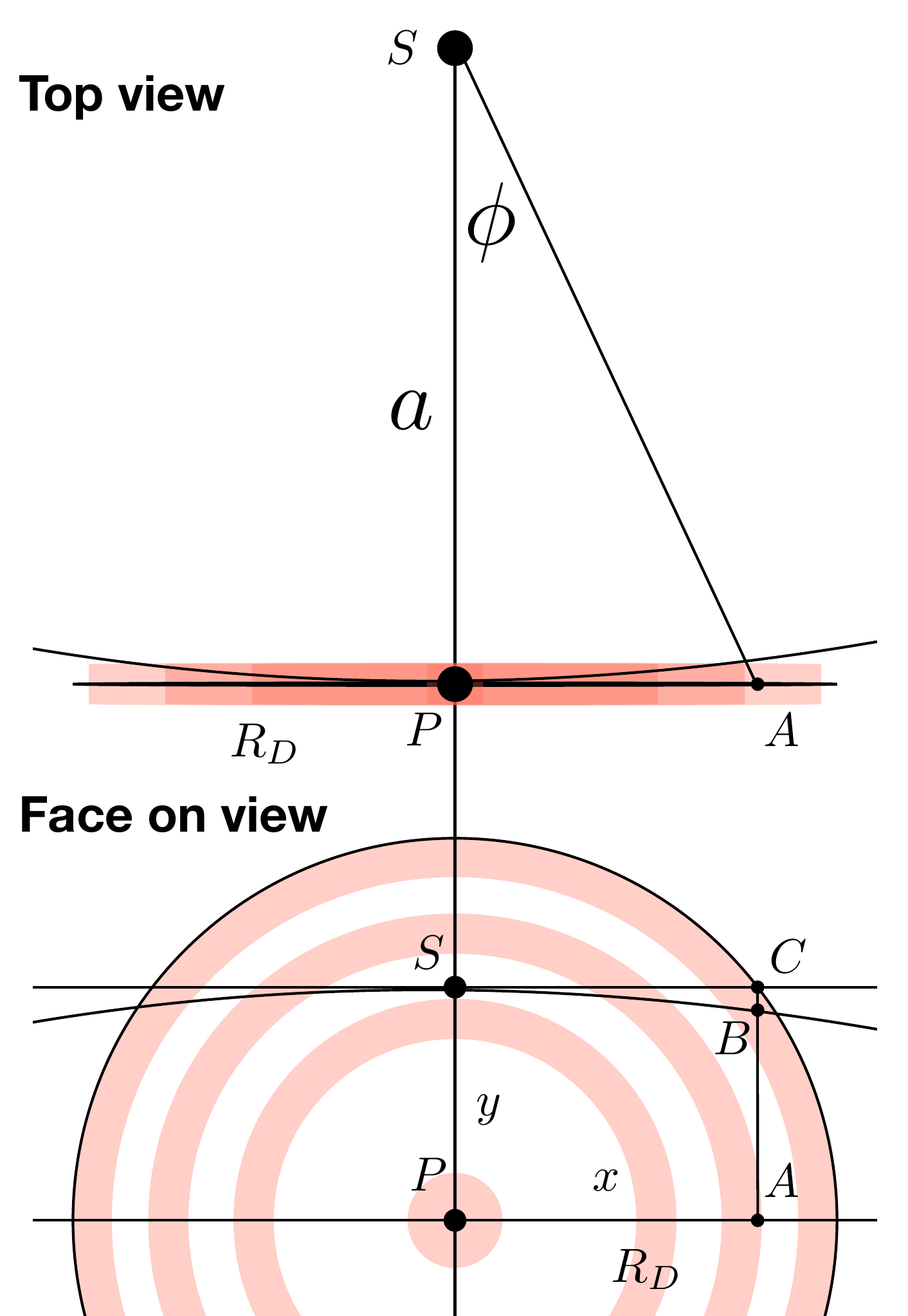}
      \caption{Sketch of geometry used in the {\tt exoring} software. There are two views of the same system, indicating the relevant distances and quantities used in the derivation of the distance $BC$.}
     \label{fig:exoringgeom}
\end{figure}

The {\tt exoring} code approximates the path of the star behind the rings with a straight line that is parallel to the x-coordinate of the ring coordinate system (line $SC$; see Figure~\ref{fig:exoringgeom}).
In reality, for nonzero impact parameters, the path of the star behind the ring system follows an ellipse (assuming a circular orbit for the eclipser around the star) whose projected semimajor axis on the sky is the radius of the orbit of the eclipser about the star, $a$.

We estimate how close the approximation of the straight line path is to the correct orbital path by taking the most extreme case scenario: that of a disc that is face-on to the observer with an impact parameter $y$.
In Figure~\ref{fig:exoringgeom}, $S$ represents the star, $P$ the centre of the disc, and $R_D$ is the radius of the disc seen face on to the observer.
The straight line path lies along the line $SC$, where $C$ is the point where the star leaves the disc.
The orbital curved path is along $SB$, and the perpendicular distance from the x-axis to the star's location is $AB=y\cos \phi$ where $\phi$ is the orbital phase angle of $P$ in its orbit around $S$.

Our goal is to calculate the distance $BC$, the distance between the idealised path and the orbital path.
From geometry, $\phi$ can be calculate from triangle $SPA$ as $\tan \phi=x/a$ which we approximate to $\phi = x/a$ and substitute into $BC$ to get $BC=yx^2/(2a^2)$.
We substitute $x^2=(R_D^2-y^2)$ to obtain:

$$BC = \frac{y(R_D^2-y^2)}{2a^2}$$

We note that $BC$ is zero both at $y=0$ and $y=R_D$, with a maximum value at $y=R_D/\surd{3}$.
We now have:
$$BC_{MAX}= \frac{r_D^3}{3\surd{3}a^2}$$

The radius of the disc and the semimajor axis $a$ are related through the radius of the Hill sphere:

$$R_H^3=a^3(1-e)^3\left ( \frac{m}{3M_*} \right )$$

...where $e$ is the eccentricity of $P$ about $S$ (taken to be zero), and $m$ and $M_*$ are the masses of $P$ and $S$ respectively.

The disc fills a fraction $\xi$ of the Hill sphere $R_H$ so we have $R_D=\xi R_H$ and substituting into $BC_{MAX}$ we get:

$$\left( \frac{BC}{R_D}\right)_{MAX} = \frac{\xi^2(1-e)^2}{3\surd{3}} \left ( \frac{m}{3M_*} \right ) ^{2/3}$$

If we take the mass ratio of the star and planet to be 0.1, this equation is now:

$$ \left( \frac{BC}{R_D}\right )_{MAX} = 0.02 \xi^2(1-e)^2 $$

This shows that for the most extreme case, the difference in the paths is 2\%, and for ring stability, typically $\xi \approx 1/3$ resulting in a difference of much less than one percent for all realistic geometries.

One can ask how valid is the approximation of a constant velocity of the star behind the rings.
The maximum velocity $v_{max}$ of the star is when the star crosses the centre line of the disk $SP$ at $\phi=0$, and the projected velocity on the sky plane $v=v_{max} \cos \phi \approx v_{max}(1 - \phi^2/2)$.
The fractional decrease in the velocity $\Delta v/v_{max}=\phi^2/2$, and in the most extreme case crossing the full diameter of the disk this is where $R_D^3 = \xi^3 a^3 (m/3M_*)$ leading to:

$$\tan \phi \approx \phi = R_D/a = \xi (m/3M_*)^{1/3}$$

Fractional decrease in velocity is therefore:

$$\Delta v / v_{max} = \frac{\xi^2}{2}\left(\frac{m}{3M_*}\right )^{2/3}$$

For $\xi=1/3$ and $m/M_*=0.1$, this is approximately 0.6\%, so the assumption of constant velocity is reasonable.

\end{appendix}
%
%
\end{document}